\documentclass[12pt,letterpaper]{article}
\pdfoutput=1

\usepackage{amsmath,amssymb,calc, amsthm}
\usepackage{graphicx}
\usepackage[nosort]{cite}
\usepackage{epsfig,psfrag}

\makeatletter
\renewcommand\section{\@startsection {section}{1}{\z@}%
                                   {-3.5ex \@plus -1ex \@minus -.2ex}
                                   {2.3ex \@plus.2ex}%
                                   {\normalfont\large\bfseries}}
\renewcommand\subsection{\@startsection{subsection}{2}{\z@}%
                                     {-3.25ex\@plus -1ex \@minus -.2ex}%
                                     {1.5ex \@plus .2ex}%
                                     {\normalfont\bfseries}}
\makeatother

\theoremstyle{plain}

\theoremstyle{definition}

\let\non\nonumber

\let\a=\alpha
\let\b=\beta
\let\h=\eta

\def\one{^{(1)}}

\newcommand{\bea}{\begin{eqnarray}}
\newcommand{\eea}{\end{eqnarray}}
\newcommand{\be}{\begin{equation}}
\newcommand{\ee}{\end{equation}}
\newcommand{\bma}{\begin{pmatrix}}
\newcommand{\ema}{\end{pmatrix}}

\newcommand{\hlf}{\frac{1}{2}}

\newcommand{\Z}{{\mathbb Z}}
\newcommand{\R}{{\mathbb R}}

\newcommand{\e}{\epsilon}

\newcommand{\dd}{\delta}
\newcommand{\com}[2]{{ \left[ #1, #2 \right] }}

\newcommand{\m}{\mu}
\newcommand{\n}{\nu}
\newcommand{\p}{\partial}

\def\com#1#2{{ \left[ #1, #2 \right] }}

\newcommand{\C}[1]{$(\ref{#1})$}

\newcommand{\diag}[1]{{\rm diag}(#1)}

\def\IZ{\relax\ifmmode\mathchoice
{\hbox{\cmss Z\kern-.4em Z}}{\hbox{\cmss Z\kern-.4em Z}}
{\lower.9pt\hbox{\cmsss Z\kern-.4em Z}} {\lower1.2pt\hbox{\cmsss
Z\kern-.4em Z}}\else{\cmss Z\kern-.4em Z}\fi}
\def\IR{\relax{\rm I\kern-.18em R}}

\def\one{{\hbox{ 1\kern-.8mm l}}}

\def\tr{{\rm tr\,}}
\def\Tr{{\rm Tr\,}}

\newlength{\bredde}
\def\slash#1{\settowidth{\bredde}{$#1$}\ifmmode\,\raisebox{.15ex}{/}
\hspace*{-\bredde} #1\else$\,\raisebox{.15ex}{/}\hspace*{-\bredde}
#1$\fi}

\newsavebox{\zzzbar}
\sbox{\zzzbar}
  {\setlength{\unitlength}{0.9em}
  \begin{picture}(0.6,0.7)
  \thinlines
  \put(0,0){\line(1,0){0.6}}
  \put(0,0.75){\line(1,0){0.575}}
  \multiput(0,0)(0.0125,0.025){30}{\rule{0.3pt}{0.3pt}}
  \multiput(0.2,0)(0.0125,0.025){30}{\rule{0.3pt}{0.3pt}}
  \put(0,0.75){\line(0,-1){0.15}}
  \put(0.015,0.75){\line(0,-1){0.1}}
  \put(0.03,0.75){\line(0,-1){0.075}}
  \put(0.045,0.75){\line(0,-1){0.05}}
  \put(0.05,0.75){\line(0,-1){0.025}}
  \put(0.6,0){\line(0,1){0.15}}
  \put(0.585,0){\line(0,1){0.1}}
  \put(0.57,0){\line(0,1){0.075}}
  \put(0.555,0){\line(0,1){0.05}}
  \put(0.55,0){\line(0,1){0.025}}
  \end{picture}}

\newcommand{\ena}{\end{eqnarray}}
\newcommand{\beqa}{\begin{eqnarray}}
\newcommand{\eeqa}{\end{eqnarray}}

\newcommand{\half}{\frac{1}{2}}

\def\bes #1\ees{\begin{split}#1\end{split}}

\renewcommand{\b}{\beta}
\newcommand{\g}{\gamma}

\newfont{\goth}{ygoth.tfm scaled 1200}                   

\def\a{\alpha}
\def\b{\beta}
\def\e{\epsilon}
\def\th{\theta}
\def\g{\gamma}
\def\h{\eta}

\def\m{\mu}
\def\n{\nu}

 \numberwithin{equation}{section}

\def\1{{(1)}}
\def\2{{(2)}}
\def\3{{(3)}}

\def\1{{\bf 1}}
\def\a{{\alpha}}

\def\M{{\mathcal M}}

\def\1{{\bf 1}}
\def\3{{\bf 3}}
\def\7{{\bf 7}}
\def\2{{\bf 2}}
\def\8{{\bf 8}}

\newcommand{\lp}{\left(}
\newcommand{\rp}{\right)}

\newcommand{\hg}{\hat g}
\newcommand{\hn}{\hat{\nabla}}

\begin{document}
\begin{titlepage}

\begin{center}

\today
\hfill         \phantom{xxx}  EFI-13-33

\vskip 2 cm {\Large \bf Domain Walls, Triples and Acceleration}
\vskip 1.25 cm {\bf  Travis Maxfield\footnote{maxfield@uchicago.edu} and Savdeep Sethi\footnote{sethi@uchicago.edu}}\non\\
\vskip 0.2 cm
 {\it Enrico Fermi Institute, University of Chicago, Chicago, IL 60637, USA}

\end{center}
\vskip 2 cm

\begin{abstract}
\baselineskip=18pt

We present a construction of domain walls in string theory. The domain walls can bridge both Minkowski and AdS string vacua. A key ingredient in the construction are novel classical Yang-Mills configurations, including instantons, which interpolate between toroidal Yang-Mills vacua. 
Our construction provides a concrete framework for the study of inflating metrics in string theory. In some cases, the accelerating space-time comes with a holographic description. The general form of the holographic dual is a field theory with parameters that vary over space-time.       


\end{abstract}

\end{titlepage}


\section{Introduction} \label{intro}

More often than not, it has been fruitful to study natural structures in string theory. For example, the study of static D-branes or NS-branes has provided interesting examples of holography. What has not appeared in any natural way in string theory are accelerating universes with either de Sitter or FRW metrics~\cite{Dasgupta:2014pma, Green:2011cn, Gautason:2012tb}.  
For this reason, our understanding of quantum gravity for the case most germane to nature is lacking. The approaches that are currently pursued typically involve analytic continuations of the AdS/CFT correspondence to gain insight into the possibility of a dS/CFT correspondence; see, for example,~\cite{Strominger:2001pn}. We will take a different approach to this problem more grounded in string theory.

The goal of this work is basically to investigate the physics of branes in string theory with one transverse non-compact dimension; these branes essentially look like particles in one spatial dimension. Gravitational back reaction is quite severe in this setting so we typically expect classical gravity to break down if we insist on static backgrounds.  However, string theory is not just classical gravity. We will describe both static and time-dependent backgrounds. Cases of backgrounds with high curvature are still amenable to a world-sheet analysis as long as the string coupling is well behaved. From a space-time rather than world-sheet perspective, stringy effects can be studied via $\alpha'$-suppressed corrections to the supergravity equations of motion. The leading $\alpha'$ corrections are best understood in the ten-dimensional heterotic or type I string space-time effective action. All the essential new ingredients that string theory brings to the table are visible in the leading four derivative interactions, which are completely known.  These include couplings that can violate the strong energy condition, along with couplings that generate gravitational sources of brane charge. 

These higher derivative couplings induce drastic effects even for large volume Calabi-Yau compactifications. For example, the gravitational correction to the heterotic Bianchi identity, 
\be\label{introbianchi}
dH = {\alpha' \over 4}\left( \tr R_+ \wedge R_+ - \tr F\wedge F\right), 
\ee 
together with a similar correction to the Einstein equations, permit us to turn on gauge field strengths along Calabi-Yau directions. Such a possibility is not allowed if we neglect the four derivative gravitational interaction appearing in~\C{introbianchi}\ and just consider supergravity couplings. The essential question for us is whether  stringy couplings permit new kinds of brane and domain wall solutions with one transverse dimension. This is basically an extension of the study of stringy effects to spaces with boundaries. Our usual intuition is that higher derivative couplings play no critical role in understanding the physics of low curvature backgrounds. As we will see, that is not the case for branes with one or two transverse directions.

Domain walls are solutions that connect distinct vacua of field theories and string theory. There has been considerable past work on domain walls in string theory, although largely focused on BPS configurations. 
The constructions of domain walls can be quite involved because the vacua of string theory are typically complicated, and often topologically distinct. Most of the constructions use some lower dimension effective supergravity theory rather than a full ten-dimensional string theory. For this reason, no intrinsically stringy phenomena have emerged from past studies of such configurations. Most such walls are constructed in a thin wall approximation reviewed, for example, in~\cite{Cvetic:1996vr}. Gravity typically propagates in the dimension transverse to the wall. In principle, a mechanism to localize gravity on the wall, like the one proposed in~\cite{Randall:1999vf}, is not prohibited in string theory but we are unaware of any past examples that concretely realize localized gravitons.\footnote{We would like to thank Andreas Karch for discussions about stringy attempts to realize localized gravitons.} Several of the constructions we will describe have the possibility of localized gravitons. The localization can happen by different mechanisms, which we will sketch below. 

The domain walls we will propose are much closer in nature to conventional string brane solutions. There are analogues of both NS-branes and D-branes. They have several nice virtues: they connect relatively simple vacua of the heterotic and type I strings which can be either Minkowski or AdS space-times. They involve no topology change or intrinsically string scale phenomena, and hence can be studied using conventional space-time techniques. The line elements for the walls are typically cosmological. In the case of walls connecting two AdS vacua, our construction provides a framework for defining and exploring the holographic description of an accelerating universe. 

We can gain some intuition about how the physics of domain walls changes because of stringy interactions by considering a one-dimensional Laplace equation, which arises in the study of electromagnetism in one spatial dimension:
\be\label{sketchsources}
\nabla^2\phi = S_{\rm supergravity} + S_{\rm stringy}. 
\ee
Let $y$ denote the coordinate for the spatial direction. If the supergravity and stringy sources appearing on the right hand side of~\C{sketchsources}\ are integrable in $y$, they provide effective charges for the potential $\phi$:
\be
Q = \int dy \, S. 
\ee
The asymptotic behavior of the potential $\phi$ for large $|y|$ is determined by the total charge, 
\be  \label{largey}\phi(y) \sim {Q\over 2} |y|.\ee 
This linear behavior reflects the familiar phenomenon that interactions grow with distance in one spatial dimension. On the one hand, supergravity sources tend to generate charges of one sign; on the other hand, stringy higher curvature interactions can contribute with the opposite sign. This suggests that a ``screening" phenomenon is possible in string theory, where a supergravity source produces a gravitational back reaction that screens its associated charge and long distance fields.  

In fact, it appears that many new phenomena are possible in string theory. We can qualitatively sketch some of the more intriguing possibilities. Assume that the wall metric takes the form, 
\be\label{firstform}
 ds^2_{\rm ||} + e^{- k |y|} dy^2, 
\ee 
for large $|y|$, where $ds^2_{\rm ||}$ is independent of $y$.  The $y$-direction is the single direction transverse to the wall. For positive $k$, the $y$ direction actually has finite volume. It spontaneously compactifies to an interval. This metric is geodesically incomplete so some additional data is needed to complete the space-time. Solutions that involve spontaneous compactification have been described in~\cite{Dudas:2000ff}. In such a case, gravity is at least formally localized on the wall. A second way that gravity can localize is for metrics of the form,
\be
 e^{- k |y|} ds^2_{\rm ||} +dy^2, 
\ee
which is along the lines described in~\cite{Randall:1999vf}. In this case, localization is most easily seen by noting that the Newton constant obtained by integrating over $y$ is finite. Lastly, we might imagine an asymptotic metric of the form, 
\be
ds^2_{\rm ||} +dy^2, 
\ee
which is not what typically comes out of a supergravity analysis because of the behavior~\C{largey}. However, the screening phenomenon described above makes this a possibility in string theory. Indeed, what we mean by a conventional domain wall is precisely this case where any scalars in the theory, along with the metric, asymptote to chosen vacuum values near infinity.  

Since we are typically dealing with non-supersymmetric backgrounds, it is difficult to find the kind of closed form beautiful solutions seen in the study of supersymmetric backgrounds. That is the price we must pay for exploring physics closer to nature. We expect a combination of analytic and numerical approaches will be needed to understand the full range of possible solutions. Our aim in this work is to lay out the basic construction with broad brush strokes. Almost every ingredient used in the construction has  associated interesting open questions, and we will spell out some of those questions later in this introduction.   

We begin in section~\ref{field}\ with a lightening review of Yang-Mills vacua on tori. This section draws heavily on~\cite{deBoer:2001px}\ and references therein. 
We need this discussion of Yang-Mills vacua because we will be coupling instanton-like Yang-Mills configurations to gravity in order to construct stringy domain walls.  We describe components of the moduli space of Yang-Mills vacua on $T^3, T^4$ and $T^5$ for the gauge groups $E_8$ and $Spin(32)$. These are the cases of prime interest for string constructions. In each case, there are distinct components in the moduli space of flat connections. It is these distinct field theory vacua that go into building simple toroidal heterotic and type I string vacua, which are disconnected by a finite energy barrier. There is no topological obstruction preventing interpolation between any two vacua.  

\begin{figure}[h]
\begin{center}
\includegraphics[width=15cm]{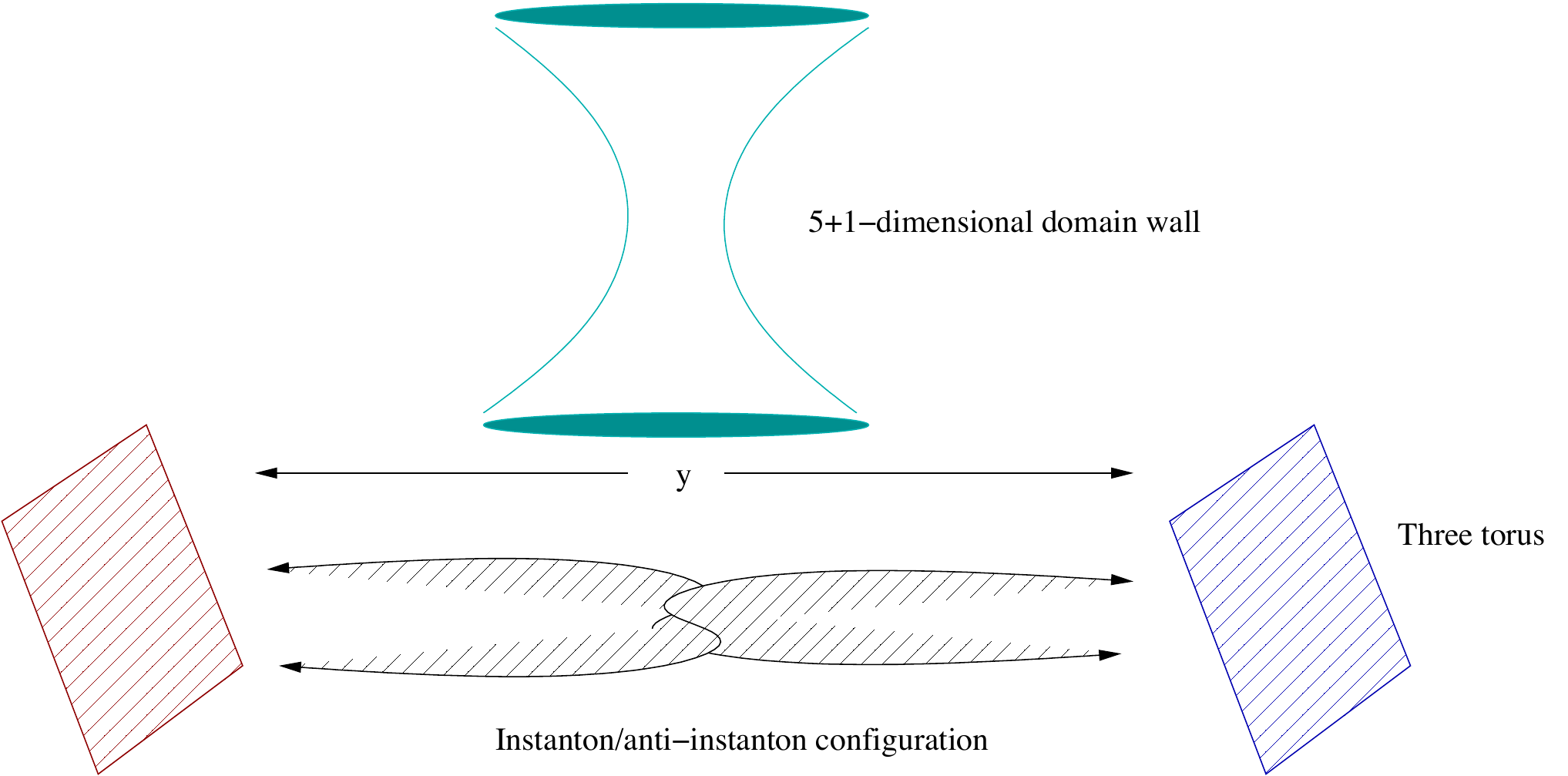}
\caption{The embedded instanton/anti-instanton pair interpolate
between fixed asymptotic CS invariants.} \label{fig:domain}
\end{center}
\end{figure}

In section~\ref{onetransversedimension}, we turn to the construction of supersymmetric brane configurations with one transverse dimension. These are essentially heterotic or type I 5-branes on $\R\times T^3$. There are two reasons for considering the supersymmetric case. First, this case provides a nice warm up where we can see the effect of stringy corrections on the usual supergravity analysis. Indeed, as we show, there are no supergravity solutions at all for this case! If a solution is to exist, it must involve the stringy higher curvature interactions. We describe the existence question that comes about from this analysis. The second reason to study the supersymmetric case is because our stringy domain walls involve a brane/anti-brane pair. The domain walls themselves are not supersymmetric, though they are built from supersymmetric ingredients. If the walls were supersymmetric, we could not hope for interesting cosmology.

Section~\ref{Minkdomain}\ contains a discussion of field theory configurations that interpolate between Yang-Mills toroidal vacua. The basic case is an instanton solution on $\R\times T^3$ which interpolates between Yang-Mills vacua labeled by their Chern-Simons $(CS)$ invariant on $T^3$.  It is important that the $CS$ invariant can be fractional for the groups $E_8$ and $Spin(32)$. There are also generalizations involving $T^4$ and $T^5$. We then describe the embedding of these instanton configurations in string theory as the basic building blocks for domain walls connecting Minkowski vacua.  The structure of these domain walls is very far from the usual picture of a thin wall. The basic setup is depicted in figure~\ref{fig:domain}.

In section~\ref{AdScase}, we turn to the question of domain walls in AdS space-times. The system we choose to study is the type I D1-D5 theory on $T^4$, or its S-dual version involving the F1-NS5 system in the heterotic string. We focus on this case as a first example largely for its simplicity.  There are many generalizations. A domain wall in this system will look like a bubble of two-dimensional space-time in the ambient $AdS_3$. 

We start by describing the distinct vacua that are possible in this case thanks to the existence of disconnected Yang-Mills vacua on $T^4$. We then discuss the distinct ways one can interpolate between these vacua. Unlike the Minkowski case, there are several choices of interpolating direction with very different boundary interpretations. For example, we could imagine interpolating along a boundary spatial direction or along the AdS radial direction. The existence and structure of interpolating Yang-Mills configurations is already a non-trivial question in this setting because the configurations must involve the bulk directions. 
Finding bulk solutions for space-time-dependent boundary gauge-field configurations requires a generalization of the kind of analysis performed in~\cite{2011JHEP...01..149K, Horowitz:2013mia}. Lastly, we discuss the general form of the holographic interpretation suggested by this construction.    

As mentioned earlier, there are many interesting questions that arise from this study in areas that span classical field theory to string theory. We will highlight a few central issues:
\begin{itemize}
\item
A classification of Yang-Mills vacua on $T^\ell$ for $\ell >3$ is in order.
\item
What can be said about instantons on $\R\times T^3$ which interpolate between different flat connections on $T^3$, and about interpolating configurations on higher tori? 
\item
Do supersymmetric NS5-branes exist on $\R\times T^3$? 
\item
What can be said about the space of gravitational solutions that interpolate between string Minkowski vacua? Specifically about possible metric singularities and asymptotic behaviors. 

\item 
What is the effective field theory description obtained by integrating out the compact directions?

\item
What can be learned about interpolating Yang-Mills configurations in the AdS case? 
\end{itemize} 
There are many more basic issues to be addressed, particularly around holography and time-dependent configurations. In terms of past work: the basic ingredients that go into these string domain walls were described in~\cite{Keurentjes:2002dc}. The existence of instanton configurations on $\R\times T^3$ which interpolate between $T^3$ vacua with fractional $CS$ invariants was suggested in~\cite{Keurentjes:1998uu, Keurentjes:2002dc}. A proof of the existence of BPS interpolating instanton configurations for specific cases appears in~\cite{Stern:2008jv}.  Past studies of domain walls in the heterotic string include~\cite{Klaput:2013nla}.

\section{Yang-Mills Vacua} \label{field}

In building domain walls in string theory, we will be using ingredients from classical field theory. There are many open questions purely in classical Yang-Mills theory that need to be addressed. We begin by describing the Yang-Mills vacuum configurations of interest to us.  

\subsection{Yang-Mills vacua on tori}

The basic ingredient needed to describe a domain wall are two vacua to which the domain wall solution asymptotes at plus and minus infinity.  Consider Yang-Mills theory with gauge group $G$ on a spatial torus $T^{\ell}$. We will restrict to topologically trivial bundles. Classical vacua correspond to a choice of flat connection satisfying $F_{mn}=0$. The solution to this condition is that the potential be pure gauge,
\be
A_{m} = -i\left(\p_{m}U\right)U^{-1},
\ee
with $U \in G$. This can be rewritten in a form, 
\be
\p_{m} U = i A_{m} U,
\ee
that can be formally integrated to give, 
\be
U({x}) = P \exp\left\{ i\int_0^{{y}} A_{m}dy^{m} \right\} U(0),
\ee
where $P$ denotes path-ordering. The integral is, of course, defined only on a path from $0$ to ${y}$, but it is path-independent (within a topological sector) because $F_{mn} = 0$.

We demand that $A$ is periodic on the torus; however, this does not imply a periodic $U$. Instead, $U$ is identified under torus translations up to a choice of holonomy $\Omega_{m}$:
\be
U( {y} + {L}^{m}) = P\exp\left\{i\int_{{y}}^{{y} + {L}^{m}} A_{n}dy^{n} \right\} U({y}) \equiv \Omega_{m}({y})U({y}).
\ee
The torus coordinates satisfy ${y} \sim {y} + {L}^{m}$. Since the connection is flat, and the integral is path-independent, the holonomies must commute: $\com{\Omega_{m}({y})}{\Omega_{n}({y})} = 0$. 

Using the fact that the gauge field is periodic, one can check that
\be
\p_{m} \left\{U^{-1}({y})U({y} + {L}^{n}) \right\} = 0.
\ee
Therefore, we write
\be \label{holonomydefinition}
U(y + L^{m}) = U(y) \omega_{m},
\ee
with $\omega_{m}$ constant commuting matrices. We can relate $\omega_{m}$ to  $\Omega_{m}(y)$ via
\be
\Omega_{m}(y) = U(y)\omega_{m}U^{-1}(y).
\ee
Fixing $U(0) = 1$, we have $\omega_{m} = \Omega_{m}(0)$. Therefore any flat connection on $T^{\ell}$ yields a set of $l$ commuting holonomies $\{\omega_{m}\}$. 

\subsection{Vacua on $T^3$}

In the case of $T^3$, a set of commuting holonomies is also sufficient for the existence of a flat, periodic connection. Namely, given a set of commuting elements $\{\omega_1,\omega_2,\omega_3\} \in G$, where $G$ is simple, connected, and simply connected, there exists a periodic flat connection with  holonomies specified by $\omega_{m}$.

The proof is constructive, and starts by using~\C{holonomydefinition} to construct $U$ along the edges of the torus cell~\cite{Keurentjes:1998uu}. This can be extended to the faces of the cell using the simply connectedness of $G$.  Finally, the definition of $U$ on the faces can be extended to the interior using the fact that $\pi_2(G) = 0$ for any simple, compact $G$.

Now that we have established the relation between commuting triples of holonomies and flat connections, we would like to describe the moduli space of such triples. The surprising feature of the moduli space is the existence of disconnected components labeled by their Chern-Simons invariant~\cite{Witten:1997bs}. These distinct components exist even though the bundles are topologically trivial. 

The trivial component of the moduli space -- the component which contains the trivial connection $A_{m} =0$ -- consists of $\omega_{m} \in T_G$, where $T_G$ is a maximal torus of $G$. The holonomies can be expressed as exponentials of the Cartan subalgebra ($CSA$) of $G$, 
\be
\omega_{m} = e^{i\a_{m}^aH^a},
\ee
where $H^a\in CSA(G)$. The flat connections can therefore be constructed using constant gauge fields. 

For groups $Spin(N\geq7)$, $G_2$, $F_4$, $E_{6,7,8}$, there also exist components of the moduli space disconnected from the trivial solution. For example, in $Spin(7)$ there is a single isolated solution disconnected from the trivial solution. Up to conjugation, this vacuum can be described explicitly by the holonomies: 
\be
\bes
\omega_1& = \diag{1, 1, 1, -1, -1, -1, -1}, \cr
\omega_2& = \diag{1, -1, -1, 1, 1, -1, -1},  \cr
\omega_3& = \diag{-1, 1, 1-, 1, -1, 1, -1}.
\ees
\ee
It is important for us that the associated flat connection cannot be realized by a constant gauge-field. In a Fourier expansion of the gauge-field on $T^3$, these new components of the moduli space involve non-trivial configurations of the massive modes; for an explicit construction of  such gauge-fields, see~\cite{Selivanov:2000kg}.

\begin{table}
\begin{center}
\begin{tabular}{|c|c|c|c|} \hline
Order of               & Maximal                   &                &    \\
the Component    & Unbroken Gauge Groups     &     Degeneracy & Dimension \\ \hline
  1            &   $E_8$                   &       1        &  24 \\
  2            &    $F_4$, $C_4$                  &       1        &  12 \\
 3            &    $G_2$                   &       2        &   6 \\
 4            &    $ A_1 $                 &       2        &   3 \\
 5            &    $\{e\}$                 &       4        &   0 \\
 6            &    $\{e\}$                 &       2        &   0 \\ \hline
\end{tabular}
\caption{The structure of the moduli space for $E_8$.} \label{table:E_8}
\end{center}
\end{table}

On $T^3$, each new component in the moduli space is uniquely labeled by its Chern--Simons ($CS$) invariant~\cite{Borel:1999bx}, which is given by
\begin{equation}
\label{CS}
 \int_{T^3} {\widetilde {CS}}(A) = \frac{1}{16 \pi^2 h} \int_{T^3} {\rm tr} \left( A dA + \frac{2}{3} A^3 \right),
\end{equation}
where $h$ is the dual Coxeter number and ${\widetilde {CS}}(A)$ denotes the Chern--Simons differential form.  The Chern--Simons 
invariant is well-defined in $\R/\Z$ and is constant over a connected
component of the moduli space. These invariants are typically rational numbers
for components that do not contain the trivial connection. 

We are mainly interested in the groups $E_8$ and $Spin(32)/\Z_2$. Tables~\ref{table:E_8}\ and~\ref{table:SO(32)}, taken from~\cite{deBoer:2001px}, summarize the structure
of the moduli spaces for $E_8$ and $Spin(32)/\Z_2$. In the latter case, for completeness
we allow a non-trivial topological choice by considering bundles both with and without vector structure.  
Note that there are $12$ distinct components for $E_8$ and $6$ for
$Spin(32)/\Z_2$. The Chern--Simons invariants for a component of order
$k$ is of the form ${n\over k}$ with $1\le n \le k$, and $n$ relatively prime
to $k$. There is exactly one component of order $k$ for each such
$n$. For example in the $E_8$ case, there are $2$ components with
$k=4$. We can distinguish these two components by their $CS$
invariants which are $1/4$ and $3/4$ (mod $\Z$), respectively.  The rank of the gauge symmetry in each non-trivial component is always lower than the trivial component. For $E_8$, table~\ref{table:E_8}\ lists the maximal unbroken gauge groups in each component.

\begin{table}
\begin{center}
\begin{tabular}{|c|c|c|c|c|} \hline
Order of       & Maximal                  &                &    & (No) \\
Component  & Unbroken Gauge Groups    &     Degeneracy & Dimension & Vector Structure\\ 
\hline
  1            &   $ D_{16}$              &       1        & 48 & VS  \\
  2            &   $ B_{12}$              &      1        & 36 & VS  \\ 
  2            &   $ D_n \times C_m, \, n+m=8$    &       2        & 24 & NVS \\   
  4            &   $ B_n \times C_m, \, n+m=5$ &      2        & 15 & NVS \\ \hline
\end{tabular}
\caption{The structure of the moduli space for $Spin(32)/\Z_2$.} \label{table:SO(32)}
\end{center}
\end{table}

\subsection{Vacua on $T^4$ and $T^5$}\label{quadquint}

Beyond $T^3$, there is currently no systematic classification of flat connections. However, there are definitely new non-trivial components in the moduli space beyond those found on lower-dimensional tori~\cite{Kac:1999gw}. These new components will play an important role later. For $Spin(32)/\Z_2$, there exists a quadruple configuration constructed as follows: take $T^4=S^1\times T^3$. On the $S^1$ factor, turn on a holonomy breaking the gauge group locally to $Spin(16)\times Spin(16)$. The group $Spin(16)$ also admits a non-trivial triple configuration with $CS$ invariant ${1\over 2}$ mod $\Z$. On $T^3$, embed $CS$ invariant ${1\over 2}$ in one $Spin(16)$ factor and $-{1\over 2}$ in the other $Spin(16)$ factor. The rank reduction for this configuration is $8$. The total $CS$ invariant is zero evaluated on any three sub-torus of $T^4$; yet the configuration cannot be deformed to the trivial connection while staying at zero energy. There are more possibilities when one includes bundles with no vector structure~\cite{deBoer:2001px}. For $G=E_8$ or $G=E_8\times E_8$, there are no non-trivial quadruples. 

A similar argument can be applied to the case of $G=Spin(16)$ to construct a quadruple on $T^4$. In turn, we can use this quadruple to build a quintuple for $G=Spin(32)$. Take $T^5=S^1\times T^4$ and choose a holonomy on the $S^1$ factor which breaks the gauge group to $Spin(16)\times Spin(16)$. Embed a quadruple in each $Spin(16)$ factor. This configuration is a quintuple with no unbroken gauge symmetry; the rank reduction is $16$. Similarly, for $G=E_8$ choose a holonomy on $S^1$ which breaks the gauge group to $Spin(16)/\Z_2$. Embedding a quadruple in this $Spin(16)$ factor results in a quintuple of $E_8$ with complete rank reduction.

\section{Branes with One Transverse Dimension} \label{onetransversedimension}

Our first goal is to study NS5-brane-like configurations on a transverse $\R \times T^3$, rather than the usual transverse $\R^4$ for a conventional NS5-brane. These objects look like particles in one spatial dimension. We will use $y$ as a coordinate for the $\R$ factor. For this section, we can equally well discuss the heterotic or type I strings. We will use an action, supersymmetry variations, and equations of motion expressed in terms of heterotic variables. 

\subsection{Equations of motion and symmetries}\label{heteom}

The string-frame heterotic action, omitting fermion couplings, takes the form:
\begin{equation}\label{hetaction}
\begin{split}
S={1\over 2\kappa^2} \int d^{10}x \sqrt{-g} \, e^{-2 \Phi}& \Big[R
+4 (\nabla \Phi)^2-{1\over 2} |{H}|^2 \cr & -
{\alpha' \over 4} \left( \tr |{F}|^2 -\tr |R_+|^2\right)   +O(\alpha'^2) \Big],
\end{split}
\end{equation}
with $\Phi$ the heterotic dilaton, and ${F}$ a Yang-Mills field strength for either $Spin(32)/\Z_2$ or $E_8\times E_8$.  The curvature $R_+$ is evaluated using the plus connection where,
\begin{equation}\label{conn}
{\Omega_\pm}= {\Omega} \pm \half {{H}} +O(\alpha'),
\end{equation}
and $\Omega$ is the usual spin connection. The definition of ${H}$ already
includes the following $O(\alpha')$ corrections,
\begin{equation}\label{aaav}
{H} = dB_2 +\frac{\alpha'}{4}   \left[ {\widetilde{ CS}}(\Omega_+) - {\widetilde{
CS}}(A)  \right] ,
\end{equation}
where $A$ is the connection on the gauge-bundle. The Bianchi identity satisfied by $H$ reads,
\be\label{bianchi}
d H = {\a' \over 4}\left\{ \mathrm{tr}\left(R_+ \wedge R_+ \right) - \mathrm{tr}\left(F \wedge F \right) \right\}. 
\ee
Our conventions for the trace are as follows: we use Hermitian generators so that terms in the action such as $\tr |F|^2$ and $\tr |R_+|^2$ are positive definite on a Riemannian space. For $Spin(32)/\Z_2$, $\tr$ is evaluated in the fundamental representation, while for $E_8$ it is $1\over 30$ times the trace in the adjoint representation. Since $R_+$ is already an antisymmetric matrix, we avoid making it imaginary -- which would be required by Hermiticity -- by instead redefining $\tr$ to be the negative of its usual form. Specifically, we use the following definitions:
\bea
\tr R_+ \wedge R_+ &=& R_{+MN} \wedge R_+^{\phantom{+}MN}, \cr
\tr|R_+|^2_{MN} &=& R_{+MPQR}R_{+N}^{\phantom{+ N} PQR},  \cr 
\tr|R_+|^2 &=& {1 \over 2} R_{+MNPQ}R_+^{\phantom{+}MNPQ}.
\eea
From the action~\C{hetaction}, we derive the following heterotic string-frame equations of motion to order $\alpha'$. These equations agree (up to signs) with the expressions appearing in~\cite{Gauntlett:2003cy, Becker:2009df}:
\bea
R+4\nabla^2\Phi-4\lp\nabla\Phi\rp^2-\hlf\left|H\right|^2 - {\alpha ' \over 4}\left( \mathrm{tr}|F|^2 - \mathrm{tr}|R_+|^2\right)  &=& 0\qquad\mathrm{(dilaton)}, \label{dilatoneom}\\
 R_{MN}+2\nabla_M\nabla_N\Phi-\hlf\left|H\right|^2_{MN}  - {\alpha'\over 4}\left( \mathrm{tr}\left|F\right|^2_{MN} - \left|R_+\right|^2_{MN} \right)&=& 0\qquad\mathrm{(Einstein)}, \label{Einstein}\\
d\lp e^{-2\Phi}\ast H\rp &=& 0\qquad\mathrm{(}B\mathrm{-field)},\label{bfield}\\
e^{2\Phi}d\left(e^{-2\Phi}\ast F\right) + A \wedge \ast F - \ast F \wedge A - F \wedge \ast H &=& 0 \qquad\mathrm{(gauge)}. \label{gaugeeom}
\eea
In the preceding equations, 
we use the convention:
\be
|\omega_{p}|^2_{MN} = {1 \over (p-1)!} \omega_{M Q_1 \ldots Q_{p-1}} \omega_{N}^{\phantom{N}Q_1 \ldots Q_{p-1}}, \quad \ |\omega_{p}|^2 = {1\over p!}\omega_{Q_1 \ldots Q_{p}} \omega^{Q_1 \ldots Q_{p}}.
\ee
Combining the dilaton equation with the trace of the Einstein equation gives the following useful relation,
\be\label{firstdilaton}
2\nabla^2\Phi-4\lp\nabla\Phi\rp^2+\left|H\right|^2 + {\alpha ' \over 4}\left( \mathrm{tr}|F|^2 - \mathrm{tr}|R_+|^2\right) =0,
\ee
which can be rewritten in the form:
\be\label{seconddilaton}
\nabla^2 e^{-2\Phi} = e^{-2\Phi} \left\{ \left|H\right|^2 + {\alpha ' \over 4}\left( \mathrm{tr}|F|^2 - \mathrm{tr}|R_+|^2\right) \right\}.
\ee
The ten-dimensional Einstein frame metric is related to the string-frame metric via:
\be\label{einstein}
  ds^2_{\rm Einstein}=e^{-{\Phi\over 2}} ds^2_{\rm string}. 
\ee
What differentiates the action, Bianchi identity and equations of motion from those of a conventional supergravity theory are the couplings that involve four derivative interactions constructed from $R_+$.

The action~\C{hetaction}\ completed with fermion couplings possesses $16$ supersymmetries. The supersymmetry variations are given by:
\bea\label{SUSY}
\delta \chi &=& F_{MN}\gamma^{MN} \epsilon, \cr
\delta \lambda &=& \left( \partial_M \Phi \gamma^M - {1 \over 12}H_{MNP}\gamma^{MNP} \right)\epsilon, \cr
\delta \psi_{M} &=&\left( \partial_M + {1\over 4} \Omega^{NP}_{-\phantom{N}M} \gamma_{NP} \right)\epsilon,
\eea
with $\chi$ the gaugino, $\lambda$ the dilation, $\psi_M$ the gravitino and $\Omega_{-}$ defined in~\C{conn}. The SUSY parameter, $\e$, is a ten-dimensional  Majorana-Weyl spinor.

\subsection{Supersymmetric brane solutions}

We are going to construct brane solutions of the form $\mathbb{R}^{5,1} \times M_4$, with the fields only depending on the coordinates of $M_4$. Let us use $m,n, \ldots$ for $4$-dimensional coordinate indices, and $a,b, \ldots$ for orthonormal frame indices. Greek indices $\mu, \nu,\ldots$ refer to the space-time coordinates for $\R^{5,1}$, while Roman indices $M,N, \ldots$ run over all $10$ space-time coordinates. After discussing the general case, we can specialize to cases like $M_4=\R\times T^3$. We will demand that our solutions are $1/2$ BPS. Understanding how the supersymmetric case works will be very helpful when we address the non-supersymmetric case needed for a domain wall solution. 

Following the discussion in~\cite{Callan:1991dj}, we look for solutions with metric, 
\be\label{susymetric}
ds^2 = ds^2_{6} + e^{2\eta_3 \Phi}\delta_{mn} dy^m dy^n,
\ee
and  field strengths,
\be\label{susyfieldstrengths}
F = \eta_1 \ast_4 F, \qquad H = 2\eta_2 \ast_4 d\Phi. 
\ee
The $\eta_i= \pm$ are signs correlated in a way that we will uncover. We have to be very careful with signs for reasons that will become clear a little later. 

The $\ast_4$ refers to the Hodge dual on $M_4$. To show that this ansatz is supersymmetric, we will need the following two identities relating to chiral spinors in $4$ dimensions
\bea\label{4diidentities}
\gamma_{mn}\epsilon_{\eta_4} &=& -{\eta_4 \over 2} \epsilon_{mnpr} \gamma^{pr} \epsilon_{\eta_4}, \cr
\gamma_{m}\epsilon_{\eta_4} &=& {\eta_4 \over 3!} \epsilon_{mnpr}\gamma^{npr}\epsilon_{\eta_4},
\eea
where $\eta_4 = \pm$ is the chirality of $\epsilon_{\eta_4}$, and $\epsilon_{mnpr}$ is the $4$-dimensional Levi-Civita tensor. The $10$-dimensional Majorana-Weyl spinor decomposes into a pair of $6$-dimensional symplectic Majorana-Weyl spinors. Such a spinor is equivalent to the tensor product of a $6$-dimensional Weyl spinor with a $4$-dimensional Weyl spinor, with a further reality condition imposed. This follows from the decomposition of the positive chirality Majorana-Weyl spinor of $Spin(9,1)$ into representations of $Spin(5,1)\times Spin(4)$, 
\be
{\bf 16} \rightarrow (\bf 4_+, \bf 2_+) \oplus (\bf 4_-, \bf 2_-). 
\ee Therefore, the $4$-dimensional identities~\C{4diidentities}\ can be lifted to identities acting on the full $10$-dimensional spinor $\e$. For the sake of simplicity, we will just write the $4$-dimensional terms.

First, let us examine the gaugino variation:
\bea
\delta \chi &=& F_{mn}\gamma^{mn}\epsilon_{\eta_4} \cr 
&=& {1\over 2}F_{mn}\left(\gamma^{mn} -{\eta_4 \over 2} \epsilon^{mn}_{\phantom{mn}pr}\gamma^{pr}\right)\epsilon_{\eta_4} \cr
&=& {1\over 2}\left( 1 -\eta_1\eta_4\right) F_{mn} \gamma^{mn} \epsilon_{\eta_4}.
\eea
To preserve supersymmetry, we therefore demand that $\eta_1 \eta_4 = 1$, i.e. that $\eta_1 = \eta_4$. In other words, a self-dual field strength annihilates a positive chirality spinor.

For the dilatino variation, we write out the components of $H$:
\be
H = 2\eta_2 \ast_4 d \Phi  = -{2\eta_2 \over 3!}\epsilon_{mnp}^{\phantom{mnp}r}\partial_r\Phi dy^m \wedge dy^n \wedge dy^p, 
\ee
which implies that, 
\be
 H_{mnp} = -2\eta_2\epsilon_{mnp}^{\phantom{mnp}r}\partial_r\Phi.
\ee
Thus,
\bea
\delta \lambda &=& \left( \partial_m\Phi \gamma^m - {1 \over 12}H_{mnp}\gamma^{mnp} \right)\epsilon_{\eta_4} \cr
&=& \left( \partial_m\Phi \gamma^m +{\eta_2 \over 6}\epsilon_{mnp}^{\phantom{mnp}r}\partial_r\Phi\gamma^{mnp}\right)\epsilon_{\eta_4} \cr
&=&  \left( 1- \eta_2\eta_4\right) \partial_m\Phi \gamma^m \epsilon_{\eta_4}.
\eea
Supersymmetry therefore requires $\eta_2 = \eta_4$.

The gravitino variation can be studied in much the same way as the gaugino variation.  Imagine we have a self-dual or anti-self-dual $4$-dimensional connection,
\be
\Omega_{\eta_5}^{ab} = { \eta_6 \over 2} \e^{abcd} \Omega_{\eta_5}^{cd}. \label{chiralityomega}
\ee 
Acting on a constant chiral spinor with chirality $\eta_4$, the gravitino variation is proportional to
\be
(1-\eta_4\eta_6) \Omega_{\eta_5}^{ab} \g_{ab} \e_{\eta_4}, 
\ee
which requires $\eta_4=\eta_6$. 
Supersymmetry only imposes~\C{chiralityomega}\ on $\Omega_{\eta_5}$, but to solve the equations of motion, we actually need to examine the curvature $2$-forms computed from $\Omega_{\eta_5}$. The duality properties for those torsional curvatures will be described later.

For the specific connection determined by the metric~\C{susymetric}, we can check this explicitly. 
We first calculate the spin connection in terms of $\Phi$,
\be
\Omega^{ab}_{\phantom{ab}m} = \eta_3\left(e^a_m e^b_n - e^a_n e^b_m\right)\partial^n\Phi,
\ee
where the orthonormal one-forms are defined by $e^a = e^a_m dx^m$.
The torsionful spin connection then takes the form:
\be
\Omega^{ab}_{\eta_5\phantom{b}m} = \eta_3\left(e^a_m e^b_n - e^a_n e^b_m\right)\partial^n\Phi - \eta_2\eta_5\epsilon^{ab}_{\phantom{ab}mn}\partial^n\Phi.
\ee
Acting on a chiral spinor:
\bea
\Omega^{ab}_{\eta_5\phantom{a}m}\gamma_{ab}\epsilon_{\eta_4} &=& \partial^n\Phi\left(2\eta_3 e^a_m e^b_n - \eta_2\eta_5 \epsilon^{ab}_{\phantom{ab}mn}\right)\gamma_{ab}\epsilon_{\eta_4} \cr
&=&  2(\eta_3 + \eta_2\eta_4\eta_5) \gamma_{mn}\partial^n\Phi \epsilon_{\eta_4}.
\eea
So, if $\eta_3 = -\eta_2\eta_4\eta_5$, which is equivalent to $\eta_3 = -\eta_5$ after accounting for the previous relations, then the gravitino variation vanishes for a constant spinor.

Summarizing, supersymmetry requires:
\be
\eta_1 = \eta_2 = \eta_4; \, \quad \eta_3 = -\eta_5.
\ee
The condition $\eta_5 = -1$ is determined by the supersymmetry variations~\C{SUSY}\ themselves rather than any particular solution, so $\eta_3 = 1$ is also independent of any solution. However, the signs of $\eta_1$ and $\eta_2$ change depending on whether the $4$-dimensional gauge connection is self- or anti-self-dual.

\subsection{Checking the Bianchi identity and equations of motion}

Satisfying the supersymmetry requirements does not guarantee a solution to the heterotic equations of motion; among the additional requirements is satisfying the non-trivial Bianchi identity~\C{bianchi}. It is going to be very useful for us to see how this background approximately solves the equations of motion in an  explicit fashion, particularly when we consider non-supersymmetric backgrounds. We will examine both the Bianchi identity and the equations of motion in this subsection.

\subsubsection{$B$-field EOM}
The easiest case to consider is the $B$-field EOM:
\be
d\left( e^{-2\Phi}\ast H \right) = - 2 \eta_1d \left( e^{-2\Phi} d\Phi \right) = \eta_1 d^2e^{-2\Phi}  = 0,
\ee
where we used that $\ast^2 = (-1)^{k(4-k)}$ when acting on a $k$-form in $4$-dimensional Riemannian space. Also, where convenient and appropriate, we will replace $\ast$ with $\ast_4$, essentially ignoring a $6$D volume form multiplying the full equation.

\subsubsection{The gauge EOM}
Next, the gauge EOM reads:
\be
D \ast F - 2d\Phi \wedge \ast F - F \wedge \ast H = 0.
\ee
However,
\be
\ast H = 2\eta_1 \ast^2 d \Phi = -2\eta_1 d\Phi,
\ee
so the gauge EOM becomes,
\be
D\ast F= 0,
\ee
where we used that $\ast F = \eta_1 F$. This is simply the Yang-Mills equation of motion, which is satisfied for an instanton connection. Since this equation is conformally invariant in $4$ dimensions, a flat space conventional instanton connection is sufficient for the conformally flat metric~\C{susymetric}.

\subsubsection{Bianchi identity}
Before analyzing the dilaton and the Einstein equations, it is convenient to look at the Bianchi identity:
\be
dH = {\alpha' \over 4}\left( \tr R_+ \wedge R_+ - \tr F\wedge F\right).
\ee
Substituting the expression~\C{susyfieldstrengths}\ for $H$ gives,
\be
2\eta_1 d\ast d \Phi = {\alpha' \over 4}\left( \tr R_+ \wedge R_+ - \tr F\wedge F\right),
\ee
but  $d\ast d\Phi = \ast \nabla^2 \Phi$, so
\be\label{laplaceondilaton}
\nabla^2 \Phi = {\alpha' \eta_1\over 8}\ast \left( \tr R_+ \wedge R_+ - \tr F\wedge F\right).
\ee
Written in terms of the flat space metric, $ds^2 = e^{2\Phi}\widehat{ds}^2$, we find a purely flat space equation Laplace equation:
\be
\hat{\nabla}^2 e^{2\Phi} =   {\alpha' \eta_1\over 4} \hat\ast \left( \tr R_+ \wedge R_+ - \tr F\wedge F  \right).
\ee
We will have to solve this equation along with the remaining equations of motion. 

\subsubsection{The dilaton and Einstein EOMs}\label{approxsolutions}

First let us assemble a collection of useful facts for this background:
\bea \label{usefulexpressionsforsimplecase}
R_{mn} &=& -g_{mn}\nabla^2 \Phi - 2\nabla_m\nabla_n\Phi  + 2 g_{mn}|\nabla\Phi|^2 - 2\nabla_m\Phi\nabla_n\Phi, \cr
R &=& -6\nabla^2 \Phi +6|\nabla\Phi|^2, \cr
|H|^2_{mn} &=& {1\over 2}H_{mpq}H_{n}^{\phantom{n}pq} = 4g_{mn}|\nabla\Phi|^2 - 4\nabla_m\Phi\nabla_n\Phi, \cr
|H|^2 &=& {1\over 6} H_{mnp}H^{mnp} = 4|\nabla\Phi|^2.
\eea
The dilaton equation~\C{firstdilaton}\ therefore reads,
\be\label{harmonicrelation}
\nabla^2\Phi = {\alpha' \over 8} \left( \tr|R_+|^2 - \tr |F|^2  \right).
\ee
Applying the relation~\C{laplaceondilaton}\ yields:
\be
\left( \tr |F|^2 - \tr|R_+|^2 \right) = \eta_1 \ast \left( \tr F\wedge F - \tr R_+ \wedge R_+ \right).
\ee
Next, the Einstein equation~\C{Einstein}\ becomes,
\be\label{simplereinstein}
g_{mn}\nabla^2\Phi = -{\alpha' \over 4} \left( \tr |F|^2_{mn} - \tr|R_+|^2_{mn} \right).
\ee
This time using the dilaton equation gives,
\be
\left( \tr |F|^2_{mn} - \tr|R_+|^2_{mn} \right) = {1 \over 2}g_{mn} \left( \tr |F|^2 - \tr|R_+|^2 \right),
\ee
which is a basic identity that must be satisfied by the gauge-field stress-energy and the metric curvature to solve the equations of motion on the nose. 

The field strength associated to a self- or anti-self-dual connection satisfies both of the properties needed above; namely, if $F = \eta_1 \ast F$ then
\be
\tr|F|^2 = \eta_1 \ast \tr F \wedge F, \qquad \tr|F|^2_{mn} = {1\over 2} g_{mn} \tr|F|^2.
\ee
Therefore, we see that if $R_+$ is self- or anti-self-dual, all of the equations of motion and the Bianchi identity will be satisfied. This requirement has been noted in~\cite{Ivanov:2009rh}, and recently in~\cite{delaOssa:2014cia}. For the supersymmetric ansatz, this requirement is almost never satisfied by the connection $\Omega_+$. The only case for which this is true is the standard embedding where $dH = 0$. This is shown in Appendix~\ref{SDtorsion}. That is the only case for which we can expect an exact solution to the heterotic equations of motion to this order in the $\alpha'$ expansion. 

In all other cases, solving the Bianchi identity does not provide an exact solution to the equations of motion; rather, the bosonic equations of motion will receive higher derivative corrections at order $(\alpha')^3$. The basic Bianchi identity~\C{bianchi}\  should not be corrected at that order, other than a shift in the definition of $\Omega_+$, but the relation~\C{susyfieldstrengths}\ between $H$ and $\Phi$ is likely to be corrected. This will modify the resulting Laplace equation~\C{laplaceondilaton}. We can see that the disagreement between Bianchi and the Einstein equation is precisely where we expect. Since $R_+$ would have been self- or anti-self-dual if $dH=0$, the violation of this chirality condition is proportional to $dH$, which is $O(\alpha')$ from~\C{bianchi}. This means we fail to solve~\C{simplereinstein}\ precisely by terms of order $(\alpha')^3$ where we expect the equations of motion to be modified.\footnote{That the equations of motion, including only the leading order $\alpha'$ terms, are not typically solved exactly is key in understanding how known heterotic vacua can be compatible with supersymmetry~\cite{Melnikov:2014ywa}. } 

Without control over the complete set of higher derivative interactions, we should therefore only expect to find approximate solutions to the equations of motion up to order $(\alpha')^2$. This is true regardless of whether we choose $M_4$ to be $\R^4$ or $\R\times T^3$. The basic intuition we employ when studying the heterotic space-time equations of motion is that the primary obstruction to completing an approximate solution to an exact solution, which defines a conformal field theory, is really the ability to solve the Bianchi identity. 

\subsection{Supergravity analysis for $\R\times T^3$}\label{SUGRAanalysis}

We can now specialize to the case where $M_4=\R\times T^3$ with coordinate $y$ for the $\R$ factor. Based on the preceding discussion, we expect the basic equation we need to solve to find a supersymmetric solution is the flat space Bianchi identity:
\be\label{basicequation}
\hat{\nabla}^2 e^{2\Phi} =   {\alpha' \eta_1\over 4} \hat\ast \left( \tr R_+ \wedge R_+ - \tr F\wedge F  \right).
\ee
The essential physics involved in solving~\C{basicequation}\ is electromagnetism in one dimension. The right hand side of~\C{basicequation}\ acts like a source of electric charge for a potential $e^{2\Phi}$. For the moment, let us assume the right hand side of~\C{basicequation}\ is integrable so we can define the total charge, 
\be
Q = {\alpha' \eta_1\over 4}  \int_{\R\times T^3} \left( \tr R_+ \wedge R_+ - \tr F\wedge F  \right). 
\ee
While the exact solution for $e^{2\Phi}$ might be complicated, the asymptotic behavior for large $|y|$ is completely determined by the charge, 
\be\label{dilatonasymptotic}
e^{2\Phi} = {Q \over 2} |y| + \ldots,
\ee
where omitted terms decay more rapidly.

Let us start by restricting to pure supergravity by setting the higher derivative interaction to zero, $R_+=0$.  Assume there exists an instanton solution satisfying, 
\be\label{instcharge}
\int_{\R\times T^3}  \, \tr (F\wedge F) = k,
\ee
where $k>0$ for self-dual connections and $k<0$ for anti-self-dual connections. For definiteness, assume positive charge $k>0$ which implies $\eta_1=+$. In section~\ref{YM4d}, we will discuss what is actually known about instantons on $M_4=\R\times T^3$ but, for the moment, we can keep the discussion general. 

A purely supergravity source therefore leads to $Q<0$ but this means no real solution for the dilaton $\Phi$ from~\C{dilatonasymptotic}. The sign in~\C{basicequation}\ is very important for this conclusion, which is why we have tracked the $\eta_i$ factors carefully, and checked the equations of motion carefully. This is very different from the cases $M_4=\R^4$ and $M_4=\R^3\times S^1$ where $e^{2\Phi}$ decays at infinity. In those cases, a gauge instanton sources a real solution for the dilaton at long distances. 

It is worth pin-pointing why we end up with such a strong constraint. Under the metric ansatz~\C{susymetric}, the dilaton equation~\C{firstdilaton}\ becomes,
\be
2\hat\nabla^2 \Phi = - e^{2\Phi} \left\{ \left|H\right|^2 + {\alpha ' \over 4}\left( \mathrm{tr}|F|^2 - \mathrm{tr}|R_+|^2\right) \right\}.
\ee
It is precisely because $e^{2\Phi}\left|H\right|^2 = 4|\hat\nabla\Phi|^2$ is asymptotically of order $1$ when $\Phi \sim |y|$ that it must be taken into account by promoting $\Phi$ to $e^{2\Phi}$. This leads to~\C{basicequation}, which gives the strong constraint. If $e^{2\Phi}\left|H\right|^2$ had been integrable, we could have treated it as a source and looked for harmonic solutions for $\Phi$ rather than $e^{2\Phi}$, which would not be subject to the very strong constraint that $Q\geq 0$.  A little surprisingly the non-supersymmetric setting, which is germane for building domain walls, will be better in this regard. 

\subsection{Inclusion of $R_+$}

Does this mean there are no supersymmetric  solutions sourced by a gauge instanton on $\R\times T^3$? That conclusion would be too hasty. The stringy $R_+$ source in~\C{basicequation}\ contributes with the right sign to potentially make the total charge $Q\geq 0$. This means we must have a curvature, rather than gauge-instanton, dominated contribution to the charge. Let us evaluate $R_+$ for the supersymmetric background to see whether it gives an integrable significant contribution. A symbolic computation code gives,
\bea
\ast \, \tr R_+ \wedge R_+ &=&  -8 \left( \left(\nabla^2 \Phi\right)^2 - \nabla_m \nabla_n\Phi \nabla^m \nabla^n \Phi\right), \cr
\tr |R_+|^2 &=& 4 \left( \nabla^2 \Phi \right)^2 +  8\nabla_m\nabla_n\Phi\nabla^m\nabla^n\Phi,
\eea
where all the derivatives are with respect to the actual metric~\C{susymetric}. Written in terms of the flat space metric, this becomes
\bea
\hat\ast \,\tr R_+ \wedge R_+ &=& -8 \left( \left( \hat{\nabla}^2 \Phi \right)^2 -  \hat{\nabla}_m \hat{\nabla}_n \Phi \hat{\nabla}^m \hat{\nabla}^n \Phi + 2 \hat{\nabla}^2\Phi |\hat{\nabla}\Phi|^2 + 4\hat{\nabla}_m\hat{\nabla}_n\Phi \hat{\nabla}^m \Phi \hat{\nabla}^n\Phi\right), \cr
\tr |R_+|^2 &=& 4 e^{-4\Phi}\left( 12 |\hat{\nabla}\Phi|^4 + \left(\hat{\nabla}^2\Phi\right)^2 + 8 \hat{\nabla}^2\Phi|\hat{\nabla}\Phi|^2 \right. \cr
&&\left.+ 2\hat{\nabla}_m\hat{\nabla}_n\Phi\hat{\nabla}^m\hat{\nabla}^n\Phi - 8\hat{\nabla}_m\hat{\nabla}_n\Phi \hat{\nabla}^m \Phi \hat{\nabla}^n\Phi\right).
\eea
To find solutions to the Bianchi identity, we need to solve the non-linear equation: 
\be\label{nonlinear}
\hat{\nabla}^2 e^{2\Phi} = {\alpha' \over 4}  \left(\hat\ast \, \tr R_+ \wedge R_+  - \hat\ast\,\tr F\wedge F  \right) =   {\alpha' \over 4}  \left( -8\left( \hat{\nabla}^2 \Phi \right)^2 +\ldots - \hat\ast\,\tr F\wedge F  \right).
\ee
This is an equation for $\Phi$ given a gauge instanton field strength $F$. For large $|y|$, $e^{2\Phi}$ behaves like~\C{dilatonasymptotic}\ for some $Q\geq 0$. This means the curvature $R_+$ vanishes at infinity, which bodes well for finding a solution.  We can gain some intuition about whether a solution is possible by ignoring the compact $T^3$ directions and rewriting the equation as a non-linear integral equation in $y$, 
\be\label{integralequation}
e^{2\Phi} \sim \int dy' |y-y'| \left\{ \hat\ast \, \tr R_+ \wedge R_+(y')  - \hat\ast\,\tr F\wedge F(y') \right\}. 
\ee
A key obstruction to existence appears to be positivity of the right hand side of~\C{integralequation}. The gauge-instanton must generate a large enough metric back-reaction, encoded in $\Phi$, to dominate the integral. Hence, the comment about curvature domination. 

The first case we might consider is simply setting the gauge instanton to zero in~\C{nonlinear}, and studying the purely gravitational response to a varying $H$-flux on $\R\times T^3$. This is already a fascinating question. One can study this question mathematically by treating~\C{nonlinear}\ as an exact equation, asking whether a $\Phi$ exists that solves~\C{nonlinear}\ on the nose. However, at order $(\alpha')^3$, we expect higher derivative corrections to the equation of motions, which will modify the precise equations we want to solve. Nevertheless, as we discussed in section~\ref{approxsolutions}, a solution to~\C{nonlinear}\ would be a strong indicator that a conformal field theory description exists.  

A few additional comments about this supersymmetric system are in order. Because the curvature vanishes at infinity, we can express the integrated Pontryagin class in terms of Chern-Simons invariants:
\be\label{pontryagin}
\int_{\R\times T^3} \tr R_+ \wedge R_+ \sim  CS(\Omega_+)\vert_{y=\infty} - CS(\Omega_+)\vert_{y=-\infty}.  
\ee
Since $\Omega_+$ is an $SU(2)$ connection with integer $CS$ invariants on $T^3$, this integral is integer when suitable normalized. We will defer a detailed discussion of Yang-Mills instantons on $\R\times T^3$ until section~\ref{Minkdomain}. However, we can already note that $SO(7)$ is the smallest gauge group for which a non-trivial vacuum component even exists. This means there is no analogue of the standard embedding, where we identify the gauge connection and $\Omega_+$, for gauge instantons that connect non-trivial vacuum components on $T^3$. On the other hand, we could consider a conventional $SU(2)$ instanton and ask about the standard embedding. This is the one case where we do expect to be able to solve the equations of motion exactly. In this case,
\be\label{standardembedding}
e^{2\Phi} = 1 + {Q\over 2}|y|. 
\ee       
For comparison: in the well studied case of $M_4=\R^4$, the dilaton behaves like $e^{2\Phi} = 1 + {Q\over r^2} $ where $r$ is the radial coordinate for $\R^4$. In that case, the source is located at $r=0$, which is at infinite distance with respect to the string frame metric. For $M_4=\R\times T^3$ the source is located at $y=0$, which is at finite distance. 

Indeed, asymptotically every solution of~\C{nonlinear}\ behaves like~\C{standardembedding}\
for some $Q\geq 0$. The string coupling is generically growing as $|y|\rightarrow\infty$.  This is not necessarily a bad thing. In the original heterotic string frame, the effective space-time Newton constant,
\be
\int d^{10}x \sqrt{g} e^{-2\Phi} R = \int d^{10}x \sqrt{\hat g} \left[ e^{2\Phi} R_{\mu\nu} g^{\mu\nu} +  {\hat{R}_{mn}}\hat{g}^{mn} \right] + \ldots,
\ee
is still finite or decreasing at large $|y|$. Gravity is therefore a good description at large $|y|$, although heterotic string perturbation theory is not useful. 

Fortunately, the heterotic action enjoys a symmetry under which the dilaton, metric and fluxes are redefined as follows,
\be\label{dualitymap}
\Phi_I = - \Phi, \qquad ds^2_I = e^{-\Phi} ds^2, \qquad H_I = H, \qquad F_I=F. 
\ee
This transformation implements the S-duality to type I so we have denoted the new fields with subscript $I$.  
The action~\C{hetaction}\ is invariant under this transformation aside from the dilaton factors accompanying the flux and gauge-field terms,
\be
\ldots - {1\over 2} |H_I|^2 - {\alpha' \over 4} e^{-\Phi} \tr|F_I|^2 + \ldots. 
\ee
In this new frame, the $\R^{5,1}$ space-time metric scales down at large $|y|$, while the $\R\times T^3$ metric still expands. However, the type I string coupling is weak so string perturbation theory is valid. 

It is a fascinating question to determine whether solutions of~\C{nonlinear}\ exist and whether the standard embedding defines a good string background, but it will take us too far from our main goal of constructing domain walls to study those questions further here.  Instead, we will turn to the construction of domain walls.

\section{Domain Walls Between Minkowski Vacua}\label{Minkdomain}

The discussion in section~\ref{onetransversedimension}\ oriented around supersymmetric brane solutions with one transverse non-compact dimension. As we showed, there are no classical supergravity solutions but there might be stringy solutions. That discussion is a nice warm up for the case of prime interest to us, which is the construction of stringy domain walls.  

\subsection{Yang-Mills instantons in four dimensions}\label{YM4d}

We will start by constructing field theory configurations that interpolate between the $T^\ell$ vacua, which we described in section~\ref{field}. As the basic case, consider Euclidean four-dimensional Yang-Mills theory with group $G$ on $\R \times T^3$. We will again use $y$ as the coordinate for the $\R$ factor, which will parametrize the direction transverse to the wall. The configurations of interest to us must interpolate between one vacuum configuration at $y\rightarrow -\infty$ to another at $y\rightarrow +\infty$. The vacua are labeled by the choice of $CS$ invariant.  If the topology of the bundles on $T^3$ at $y\rightarrow \pm\infty$ are identical, there is no obstruction to building a finite energy configuration that interpolates between the vacua. By finite energy, we mean finite Euclidean Yang-Mills action:
\be S_{D=4} = {1\over (g_4)^2}  \int dy \int_{T^3} \Tr (F \wedge \ast F).  \label{energyfunctional}
\ee
Here we use the notation $\Tr$ to denote the trace in an arbitrary representation of the gauge group $G$, as opposed to $\tr$ defined in section~\ref{heteom}.
Such a configuration is a kind of Yang-Mills instanton with the property that the instanton charge, 
\be\label{pont}
{1\over 8\pi^2 N_R} \int dy \int_{T^3} \, \Tr (F\wedge F) = CS\vert_{y=\infty} - CS\vert_{y=-\infty} ,
\ee  
is typically fractional. An integer change in $CS$ invariant would correspond to a conventional instanton on $\R \times T^3$. The constant $N_R$ appearing in~\C{pont}\ depends on which representation of $G$ is considered; for example, $N_R$ is twice the dual coxeter number when $R$ is the adjoint representation. For conventional instanton configurations, $N_R$ is chosen so the smallest possible instanton charge is $1$. Let us set,
\be\label{instcharge}
\int dy \int_{T^3} \, \Tr (F\wedge F) = k,
\ee
where $k>0$ for self-dual connections and $k<0$ for anti-self-dual connections.

The existence of finite energy interpolating configurations is essentially clear. One could build such a configuration and let it relax to some minimum energy, which would be an extremum of the energy functional~\C{energyfunctional}. 
The really interesting question is not whether finite energy interpolating solutions exist but whether BPS instanton configurations exist, which interpolate between triple vacua. Such configurations satisfy the minimum energy condition:
\be\label{selfdual}
F = \pm \ast F.  
\ee
Stern has provided an affirmative answer to the existence question for the case of groups with a single non-trivial vacuum component~\cite{Stern:2008jv, stern-geometry}; this result applies, for example, to the groups $G=G_2, \, Spin(N\geq7)$. For earlier discussions of such instanton solutions, see~\cite{Keurentjes:1998uu, Selivanov:2000kg, Keurentjes:2002dc}.  We suspect that this nice result can be strengthened to more general cases. For our purpose of building stringy domain walls, it is not essential to have BPS configurations. We really only need finite energy configurations. However, finding gravity solutions is much easier under the assumption that BPS configurations exist so we will make that quite reasonable assumption in our subsequent discussion.  

If we want to study cases where this assumption is proven, we could compactify the $E_8\times E_8$ heterotic string on $T^3$ and turn on a triple configuration breaking the gauge group to a group where a BPS configuration is known to exist. For example,  the order $3$ component listed in table~\ref{table:E_8}\  includes $G_2$ as a maximal unbroken gauge group. We can then build BPS configurations in the effective $7$-dimensional theory obtained by compactifying the $E_8\times E_8$ string on $T^3$ with opposite $CS$ invariants embedded in each $E_8$ factor. 

The most important open issue concerning these instantons is the moduli space of solutions. We expect such a moduli space to include an $\R$ factor parametrizing the position of the instanton in the $y$-direction. In fact, it is reasonable to expect an $\R\times T^3$ factor specifying the position of the instanton in all four dimensions. The interesting, currently unresolved, question concerns additional moduli. Specifically, whether there is a scale modulus for the instanton configuration, analogous to the scale modulus that exists for instantons on $\R^4$ as a consequence of the broken conformal invariance. 

A similar question can be asked for conventional instantons on $\R\times T^3$, which interpolate between integer $CS$ invariants. 
It is known that a charge $1$ instanton solution exists on $\R\times T^3$ for $G=SU(2)$, unlike the case of $T^4$~\cite{vanBaal:1995eh, taubes-jdg, Ford:2000zt}. However, even for conventional instantons on $\R^{4-n}\times T^n$ with $n\geq 1$ the existence of a scale modulus is unclear. This is important for string theory applications since new light degrees of freedom can appear if an instanton can shrink to zero size~\cite{Witten:1995gx}. 

For $\R^3\times S^1$ and $\R^2\times T^2$, the associated  defect or impurity gauge theories given in~\cite{Sethi:1997zza, Kapustin:1998pb}\ possess both Higgs and Coulomb branches. The Higgs branches encode the moduli space of instantons via Nahm or Hitchin equations with sources, but the existence of a Coulomb branch suggests that the instantons can shrink to zero size. This picture does not extend to $\R\times T^3$ in any obvious way. Indeed, there are indications that the scale might be set by the choice of flat connections at infinity for this case of primary interest to us~\cite{vanBaal:1998hm, vanBaal:1995eh}. We will proceed under the assumption of smooth finite size BPS solutions satisfying~\C{selfdual}, leaving a more detailed study of the moduli space and its possible singularities for subsequent work.  

\subsection{Interpolating Yang-Mills configurations in five and six dimensions}\label{quadquint}

We can extend these interpolating solutions to higher dimension. Let us consider Euclidean $5$-dimensional Yang-Mills theory on $\R\times S^1\times T^3$ with gauge coupling $g_5$. Let the $S^1$ have radius $L_1$, while the gauge coupling $(g_5)^2$ has dimensions of length. Imagine constructing a Yang-Mills configuration that interpolates between a $Spin(32)$ quadruple on $S^1\times T^3$ at $y=+\infty$ and a connection in the trivial component of the moduli space at $y=-\infty$. While there is no obvious five-dimensional analogue of the condition~\C{selfdual}, we can use the construction of the quadruple described in section~\ref{quadquint}\ to build an interpolating solution. 

Choose a holonomy around $S^1$ which breaks $Spin(32)$ to $Spin(16)\times Spin(16)$. The quadruple construction involves embedding $CS$ invariant $+{1\over 2}$ in one factor and $-{1\over 2}$ in the other factor. To unwind this configuration, embed an instanton of the type described in section~\ref{YM4d}\ in the first factor and an anti-instanton in the second factor.  This configuration is completely independent of the $S^1$ coordinate. Exciting the $S^1$ coordinate can only increase the energy of this field configuration.  It is therefore a local minimum of the Yang-Mills action which interpolates between the two components of the moduli space. We do not know whether this is a global minimum in the space of field configurations that interpolate between the quadruple and the trivial component, but a local minimum suffices for our purposes.  We can estimate the action for this field configuration:
\be
S_{D=5}  = {(2\pi L_1)(8\pi^2 N_R) \over (g_5)^2}.
\ee
In a similar way, we can consider $6$-dimensional Yang-Mills theory on $\R\times S^1\times S^1\times T^3$ with gauge coupling $g_6$. The two circles have radii $L_1$ and $L_2$. We can unwind the quintuple of $Spin(32)$ using the same construction above. The quintuple is constructed by embedding a quadruple in each factor of a $Spin(16)\times Spin(16)$ subgroup. We can again estimate the action for the interpolating configuration,
\be
S_{D=6}  = {2 (2\pi L_1)(2\pi L_2) (8\pi^2 N_R) \over (g_6)^2}.
\ee
For the quintuple of $E_8$, the action of the interpolating configuration is smaller by a factor of 2. 

\subsection{The setup for a string theory domain wall}

Now we would like to use the instantons described in section~\ref{YM4d}\ to build domain walls in string theory. Yang-Mills configurations embed naturally in the type I and heterotic strings. The $4$-dimensional Yang-Mills instantons define NS5-brane-like configurations in string theory as we described in section~\ref{onetransversedimension}. The world-volume of the NS5-branes fill out the $6$ space-time dimensions transverse to the instanton configuration. We will start by focusing on this case, which is the basic building block, rather than higher-dimensional Yang-Mills configurations. 

To be honest domain walls in string theory, the instantons must bridge honest string vacua. Consider the $E_8\times E_8$ string on $T^3$. We can build a vacuum by embedding equal but opposite $CS$ invariants in the two $E_8$ factors so that $H$, defined by
\be\label{defH}
H = dB + {\a' \over 4}\left\{CS\left(\Omega_+\right) - CS\left(A_1\right)  - CS\left(A_2\right)\right\},
\ee
can be set to zero. The $E_8\times E_8$ gauge-fields are denoted $(A_1, A_2)$ in~\C{defH}. 
For a flat torus metric, $\Omega_+=0$. The dilaton is constant. The shape and size of the $T^3$ are arbitrary. These are the heterotic vacua described in~\cite{deBoer:2001px}.  

To interpolate from one triple vacuum, characterized by $CS(A_1)$, to another triple vacuum, we can embed an instanton of the type described in section~\ref{YM4d}\ in one $E_8$ factor, and an anti-instanton in the other $E_8$ factor. For definiteness, let us choose:
\be
F(A_1) = \ast F(A_1), \qquad F(A_2) = -\ast F(A_2). 
\ee
We will denote $F(A_i)$ by $F_i$ for convenience. The instanton charge~\C{instcharge}\ is $k>0$ for $F_1$ and $-k$ for $F_2$. 

The basic structure looks like a brane and an anti-brane, which is quite different from what we might expect from a thin-wall approximation. We will refer to this setup interchangeably as a brane/anti-brane or instanton/anti-instanton configuration. There is one caveat with this terminology: namely, the brane and anti-brane cannot easily annihilate! Each is associated to a distinct gauge group. The only possible annihilation channel involves either the instanton or anti-instanton shrinking to zero size and traversing the heterotic M-theory interval. Whether such a process is even possible is unclear. In the absence of gravity, this is a static Yang-Mills configuration since the two $E_8$ factors do not communicate. 

Let us use $y^m$ as an indexed coordinate for $\R\times T^3$ with $(y^1, \ldots, y^4) = (y, \th^1,\th^2, \th^3)$,  and  $x^\mu$ with $\mu=0,\ldots,5$ as coordinates for the $6$ transverse directions.  Each instanton has at least one field theoretic normalizable zero mode corresponding to the location of the instanton in the $y$-direction. We can label the positions of the instanton and anti-instanton by $(y_1, y_2)$. In an effective field theory approach, these normalizable modes give rise to $6$-dimensional scalar fields $(y_1(x), y_2(x))$. 

Our intuition about branes and anti-branes suggests that this configuration should not remain static when coupled to gravity. If we have two mutually BPS NS5-branes then the gravitational interaction between them would cancel against $B_2$-exchange, permitting a static configuration. For a brane/anti-brane, the gravitational interaction is unchanged but the $B_2$-exchange force now adds rather than cancels that interaction. As an ansatz in formulating a time-dependent string background, we will only allow time-dependence in the internal directions that is a consequence of time-dependence for $(y_1, y_2)$. NS5-branes are heavy objects at weak string coupling so we might suspect that when the brane/anti-brane are very far separated, we can treat the background as static to a first approximation.   


Let us assemble the data we want to use to construct a string background: we will take an agnostic view on the existence of additional moduli for the instantons. That is a fascinating question of classical field theory, but our focus is on the dynamics of the two fundamental scalars $(y_1, y_2)$.  In accord with our discussion of section~\ref{YM4d}, we will very reasonably assume there exist BPS self-dual $E_8$ instantons on $\mathbb{R}\times T^3$, which interpolate between the chosen triple vacua. 

This setup is depicted in figure~\ref{fig:domain}\ of the introduction. The brane/anti-brane configuration breaks all supersymmetry; however, the breaking is mild. If we separate the pair by a large distance, we might expect that the gravitational field configuration should be well approximated by a BPS configuration. As we will see, however, there are significant differences in the gravity solution because of the presence of both a brane and an anti-brane, even when they are very far separated. The intuitive reason for this long range coupling is that interactions in one spatial dimension (like the Coulomb interaction described in the introduction) grow rather than decay with distance.   

Based on the physical picture discussed above, let us make an ansatz for the full ten-dimensional string background. The internal string-frame metric components depend only on $(y, \th^i)$, in accord with a domain wall picture. Ideally, we seek a gravitational solution capturing the response to the Yang-Mills stress-energy in which the $T^3$ metric and dilaton asymptote to vacuum values at large $|y|$. Whether that is possible is not clear, but it is a natural behavior for a domain wall configuration. A reasonable metric ansatz takes the form,
\bea
ds^2  &=& e^{2w_1}  {\widehat{ds^2}}, \label{definehat} \\ 
&=& e^{2w_1} \left\{ ds^2_{\rm space-time}(x) +e^{2w_2} \left( dy^2 + e^{2w_3} L^2 \dd_{ij} d\th^i d\th^j \right) \right\}, \label{metricansatz}
\eea
and depends on three scalar functions $(w_1, w_2, w_3)$ of the coordinates $y^m$. The torus is taken to be square with sides of length $L$.  For the moment, we are ignoring any time-dependence in the internal metric. This means we might encounter a potential for the scalars $(y_1, y_2)$ in the effective $6$-dimensional theory, which we can worry about later. 

This ansatz could certainly be more complicated. For example, we can take a more general metric for $\R\times T^3$, or we might imagine a warp factor for the space-time metric that depends on both $(y,\th^i)$ and $t$. The other natural modification is to take different warp factors for the spatial and time components of the space-time metric. For the moment, we will start with~\C{metricansatz}; if the physics suggests a more general metric, we can revisit this ansatz.  

For the space-time metric, we will assume something reasonable: either a maximally symmetric space-time with cosmological constant $\Lambda$, or an FLRW metric,
\be
ds^2_{\rm space-time}(x) = -dt^2 + a^2(t) h_{ij} dx^i dx^j, 
\ee
where the spatial part of the metric takes the form, 
\be
h_{ij}= \delta_{ij} + k {x_i x_j \over 1-kx^2}. 
\ee
Any cosmology is generated only in response to the Yang-Mills background. Along with the metric, we also have an $H$-field and a dilaton $\Phi$ that need to be specified to describe the string background. 

Lastly, a comment on the $\alpha'$ expansion is in order. We will be using the $\alpha'$ expansion in the following way: first, the equations of motion~\C{dilatoneom}--\C{gaugeeom}\ themselves receive corrections at order $(\alpha')^3$ and above, which we are neglecting. For that reason, we can only expect to solve the Bianchi identity and the equations of motion to order $(\alpha')^2$, as we already saw in our discussion of the supersymmetric case. There might be special situations akin to the standard embedding of section~\ref{onetransversedimension}\ in which exact solutions to the equations of motion can be found. 


The stress-energy of the Yang-Mills field is already $O(\alpha')$.  These Yang-Mills fields depend non-trivially on all coordinates $y^a$. They vanish as $|y|\rightarrow 0$. We ideally want to solve the string equations of motion in terms of the Yang-Mills configuration, which is input data. Since the gravitational response to the Yang-Mills stress-energy is $O(\alpha')$, at leading order we can ignore the gauge-fields entirely and start with a string-frame metric of the form, 
\be\label{leadingmetric}
ds^2 = \eta_{\m\n} dx^\m dx^\n + \left( dy^2 + L^2 \dd_{ij} d\th^i d\th^j \right), 
\ee
which is a product of $6$-dimensional Minkowski space-time with a flat ``internal'' metric for $\R\times T^3$. This should be the $\alpha' \rightarrow 0$ limit of~\C{metricansatz}. The dilaton is constant. 
The worry is that the gravitational response to this stress-energy in a problem with effectively one transverse dimension can be large, but this is a reasonable way to proceed. 

This approach is different from the supersymmetric case studied in section~\ref{onetransversedimension}, where solving the supersymmetry variations provided a way to construct approximate, and in one case exact, solutions to the equations of motion.\footnote{The one case where an exact solution exists is the usual symmetric NS5-brane with $dH=0$ on the nose.} That is the magic of supersymmetry, but also the reason why supersymmetric backgrounds do not give realistic cosmologies. On the other hand, non-supersymmetric solutions are simply much harder to construct. The basic reason we can hope a solution exists in string theory is because the basic building block for this string theory domain wall is the quite beautiful purely field theoretic domain wall.  

\subsection{The Bianchi identity and equations of motion}
\subsubsection{The gauge-field equation of motion}\label{gaugeeommink}

Let us take a look at the equations of motion. 
The basic input data is a pair of finite energy interpolating gauge connections $(A_1, A_2)$. For a general metric of the form~\C{metricansatz}, the gauge field equation of motion~\C{gaugeeom}\ does {\it not} reduce to a standard Lorentz invariant $4$-dimensional Yang-Mills equation. Rather, it reduces to an effective problem on $\R \times T^3$ with additional inserted factors of $\Phi$ and the metric warp factors. We will meet this same issue in an unavoidable way when we examine domain walls in AdS space-time.  Here we will assume that the leading metric is~\C{leadingmetric}\ and that the dilaton is constant to leading order.

The leading order terms in~\C{gaugeeom}\ then amount to the requirement that the gauge-fields satisfy the Yang-Mills equations. This is true for any minimal energy instanton. This is really the only constraint we need to satisfy when considering the gravitational response at leading order in $\alpha'$ since the stress-energy produced by the gauge-fields is already $O(\alpha')$. At the next order, we encounter a more interesting constraint:
\be\label{dilrelation}
d\ast F^{(1)} + \left[ A^{(1)}, \ast F \right]  + \left[ A^{(0)}, \ast F^{(1)} \right] - F^{(0)}\wedge\ast H - 2 d\Phi \wedge \ast F^{(0)} = 0. 
\ee 
The notation $A^{(i)}$ refers to terms of order $(\alpha')^i$ in an expansion of the gauge-field,
\be
A=A^{(0)} + A^{(1)} + \ldots,
\ee 
where the leading order term, $A^{(0)}$, is the instanton or anti-instanton connection. Equation~\C{dilrelation}\ should be viewed as determining $A^{(1)}$ in terms of the known data $A^{(0)}$ and the $O(\alpha')$ solutions for $(H, \Phi, w_1, w_2, w_3)$, which we have yet to determine.

 \subsubsection{The Bianchi identity}
 
 The physics changes quite significantly from the supersymmetric case considered in section~\ref{onetransversedimension}, and this is most immediately apparent in the heterotic Bianchi identity:  
 \be\label{nonsusybianchi}
d H = - {\a' \over 4}\left\{  \mathrm{tr}\left(F _1\wedge F_1 \right)+ \mathrm{tr}\left(F_2 \wedge F_2 \right)  -  \mathrm{tr}\left(R_+ \wedge R_+ \right)\right\}. 
\ee
Let us momentarily ignore the $4$ derivative $R_+$ coupling in~\C{bianchi}.  The heterotic Bianchi identity~\C{bianchi}\
requires a non-vanishing $H$-field, despite the fact that neither asymptotic vacuum possesses a non-vanishing $H$-field. The right hand side of~\C{nonsusybianchi}\ is trivial in cohomology since on $\R\times T^3$, every $4$-form is trivial, but not point-wise zero. 

The obstruction to solving~\C{nonsusybianchi}\ with an $H$ decaying to zero as $|y|\rightarrow \infty$ is that the right hand side of~\C{nonsusybianchi}\ integrate to zero. In our case, this obstruction vanishes because the Yang-Mills instanton and anti-instanton charges cancel. Therefore the $H$ required by the Bianchi identity is completely determined by the Yang-Mills background; specifically, the right hand side of~\C{nonsusybianchi}\ is proportional to the volume form of $\R\times T^3$:
$$ \alpha' h(y^a) dy d\th_1 d\th_2 d\th_3. $$
The function, $h(y^a)$, is globally defined on $\R\times T^3$ and decaying sufficiently rapidly as $|y|\rightarrow \infty$ to ensure a finite Yang-Mills energy.  This implies that $h$ is at worst, 
\be
h \sim o\left({1\over |y|}\right),
\ee
as $|y|\rightarrow \infty$. 

Let us take a slight detour and revisit the case of a 't Hooft charge $1$ instanton on $\R^4$~\cite{tHooft:1976fv}. In that case, we see that the worst case decay of $|F|^2$ needed for finite energy only requires
\be
|F|^2 \sim o\left({1\over |y|^4}\right), 
\ee  
where $y^a$ momentarily denote coordinates for $\R^4$.  However, the actual charge $1$ solution has a field strength such that $|F|^2 \sim {1\over |y|^{8}}$, which is much more localized than the worst case decay would demand.

Returning to our instantons on $\R\times T^3$, we see there is good reason to expect that $|h| \sim |F|^2$ might decay more rapidly than the worst case.  For the fractional instantons of interest to us, the asymptotic behavior of the instanton field strength has yet to be determined. However, for conventional $SU(2)$ instantons on  $\R\times T^3$, Charbonneau has provided a set of decay estimates~\cite{Charbonneauthesis, Charbonneau:2004ak}. These estimates depend on the choice of flat connection at infinity. Aside from a discrete set of choices, the decay is faster than polynomial. Namely,
\be
|F| \sim o\left({1\over |y|^m}\right),
\ee
for any $m$ as $|y|\rightarrow \infty$. Included in the discrete set for which this is not true is the trivial connection (up to gauge transformation). In this case, the decay estimate shows,\footnote{This decay estimate does not mean that an instanton exists with $|F| \sim {1\over |y|}$. An example is known, found on p.58 of~\cite{Charbonneauthesis}, for which $|F| \sim {1\over |y|^2}$. We want to thank Benoit Charbonneau for correspondence on aspects of his work.}
\be
|F| \sim o\left({1\over |y|}\right).
\ee
This is interesting and a little surprising. It suggests that the string solution can change quite significantly if one tunes the flat connection at infinity to special values. In particular, the conventional heterotic vacuum on $T^3$ might play a distinguished role. It is going to be very interesting to understand how the instantons in our case behave more precisely, but for now, it seems very plausible that we have faster than polynomial decay of the field strength for generic choices of triple vacua. 

How does the $R_+$ coupling change these conclusions? As long as we insist on a connection $\Omega_+$ with  zero instanton number, there is no change in the conclusion that $H$ decays nicely as $|y|\rightarrow \infty$. Indeed under such an assumption we can ignore the  $H$ terms in $\Omega_+$ since $H$ is formally $O(\alpha')$, and impose the requirement of zero instanton charge on the spin connection for $\R\times T^3$.  This is a mild constraint on the warp factors $(w_1, w_2, w_3)$ appearing in~\C{metricansatz}. 


The upshot of our study of the heterotic bianchi identity is that $H$, which is $O(\alpha')$,  decays rapidly as $|y|\rightarrow\infty$ as long as
\be \int  \mathrm{tr}\left(R_+ \wedge R_+ \right) =0. \label{toponR+}
\ee This is very good news by comparison with the supersymmetric case. As discussed in section~\ref{SUGRAanalysis}, it was precisely because $|H|^2$ was not integrable that we ran into problems with the dilaton equation at the level of supergravity. There is at least a chance we can avoid that issue for these domain walls.

\subsubsection{The space-time Einstein equations}

To get a handle on $w_1$ and the space-time metric, it is useful to look at the space-time components of the Einstein equations~\C{Einstein}. The presence of a dilaton has removed many sources of stress-energy from~\C{Einstein}, which would be present in a generic gravity theory. This has quite dramatic implications, which can be seen as follows:  the Ricci curvature for the warped metric~\C{metricansatz}\ is nicely expressible in terms of the Ricci curvature for the hatted product metric~\C{definehat}\ along with derivatives of the warp factor: 
\be\label{truncatedRmn}
R_{MN} = \hat{R}_{MN} - \hat{g}_{MN} \hat\nabla^2 w_1 +8 \left( \hat\nabla_M w_1 \hat\nabla_N w_1 - \hat\nabla_M\hat\nabla_N w_1 - \hat{g}_{MN} |\hat\nabla w_1|^2 \right). 
\ee 
Restricting to space-time directions gives,
\be\label{spacetimetruncatedRmn}
R_{\mu\nu} = \hat{R}_{\mu\nu} - \hat{g}_{\mu\nu} \hat\nabla^2 w_1 -8 \hat{g}_{\mu\nu} |\hat\nabla w_1|^2. 
\ee 
The $w_1$ terms of~\C{spacetimetruncatedRmn}\ potentially have $y^m$-dependence proportional to $\hat{g}_{\mu\nu}$, which must be canceled by stress-energy sources since $\hat{R}_{\mu\nu}$ appearing in~\C{spacetimetruncatedRmn}\ is independent of the coordinates $(y, \th^1, \th^2,\th^3)$. 
On the other hand the stress-energy sources for~\C{Einstein}\ take the form,
\be\label{sources}
2\nabla_\mu\nabla_\nu\Phi-\hlf\left|H\right|^2_{\mu\nu}  - {\alpha'\over 4}\left( \mathrm{tr}\left|F\right|^2_{\mu\nu} - \left|R_+\right|^2_{\mu\nu} \right). 
\ee
For the moment, let us assume no time-dependence for the internal fields. If space-time is not $3$-dimensional,  which is the case for our metric~\C{metricansatz}, then there is no $H$-field source compatible with the symmetries of space-time that can generate a term proportional to $\hat{g}_{\mu\nu}$. The same is true for the source  $\mathrm{tr}\left|F\right|^2_{\mu\nu}$. The only source of stress-energy that has the right form comes from the dilaton which, using equation~\C{2covder}, gives a contribution:
\be
\hat{g}_{\m\n} \hat{\nabla}^P w_1 \hat{\nabla}_P\Phi.
\ee

Now let us examine some of the possibilities. If $w_1$ is constant then there is no cosmological constant for $\hat{R}_{\mu\nu}$. Space-time is not warped over the $y$-direction at all and there is no constraint on the dilaton. If $w_1$ is not constant, there can be a cosmological constant for the space-time metric with 
\be\label{maximally}
\hat{R}_{\mu\nu} = {1\over 2} \Lambda\hat g_{\mu\nu}. 
\ee
That $\Lambda$ is precisely constant is a strong condition. Ignoring the $\left|R_+\right|^2_{\mu\nu}$ term, the cosmological constant takes the form
\bea\label{definelambda}
{1\over 2}\Lambda &=& \hat\nabla^2 w_1 + 8 |\hat\nabla w_1|^2 - \hat{g}_{\m\n} \hat{\nabla}^m w_1 \hat{\nabla}_m\Phi \cr &=& {1\over 8}e^{-8w_1}\hat\nabla^2 e^{8w_1} - \hat{g}_{\m\n} \hat{\nabla}^m w_1 \hat{\nabla}_m\Phi. 
\eea
The $\left|R_+\right|^2_{\mu\nu}$ term modifies this in a way determined by~\C{curvaturesquared}. If we ignore the dilaton for a moment then  $e^{8w_1}$ must be an eigenfunction of the internal Laplacian $\hat\nabla^2$. Any normalizable eigenfunction of $\hat\nabla^2$ would have a negative definite eigenvalue giving an AdS space-time. We typically expect the scale of the cosmological constant to be set by the characteristic size $L$ of the torus. 

On the other hand, we could imagine a non-normalizable solution for $w_1$. We can use formulae for  warped metrics to write $e^{-8w_1}\hat\nabla^2 e^{8w_1}$ in terms of flat space derivatives for the metric~\C{leadingmetric}. These formulae are assembled in Appendix~\ref{usefulformulae}\ for convenience. We arrive at an expression in term of the flat space metric~\C{leadingmetric}, 
\be
{1\over 8}e^{-8w_1}\hat\nabla^2 e^{8w_1} = \sum_m \left( e^{-2w_2 -2w_3} \left( \p_m  + \p_m w_3 \right) + 2e^{-2w_2} \p_m w_2  + 8 \p_m w_1 \right) \p_m w_1, 
\ee
where the subscript $m$ refers to the internal coordinates $y^m$. To simplify things, let us take $w_2=w_3=0$. In this case, $w_1=ky$ for some constant $k$ is an eigenfunction that can give a positive $\Lambda$. For this solution, we also need $\Phi$ either linear in $y$ or independent of $y$ to ensure a constant $\Lambda$. This will certainly create tension with other equations of motion, but it is still an interesting possibility to ponder here as a way to generate acceleration. Unavoidably in such a scenario, the scale factor for space-time crunches either as $y \rightarrow  +\infty$ or $y \rightarrow  -\infty$. 

More generally, we see that solving the space-time Einstein equations gives a strong condition on the dilaton:
\be\label{determinedilaton}
 \hat{\nabla}^m w_1 \hat{\nabla}_m\Phi = \left( {1\over 8}e^{-8w_1}\hat\nabla^2 e^{8w_1}  - {1\over 2}\Lambda \right).
\ee
This equation is further modified by the $\left|R_+\right|^2_{\mu\nu}$ term, but fortunately using~\C{curvaturesquared}\ we note that the form of this correction is also proportional to either $\hat{g}_{\mu\nu}$ or $\hat{R}_{\m\n}$, and so it can be accommodated in~\C{determinedilaton}. In summary for the static case: either $w_1$ is constant or the dilaton is related to $w_1$ and the desired cosmological constant via~\C{determinedilaton}. 

We can also immediately see how to construct more general cosmologies. Without some time-dependence for the internal fields, the space-time metric is maximally symmetric with a curvature satisfying~\C{maximally}. We want to restrict the time-dependence to be as simple as possible so we  permit the zero modes $(y_1, y_2)$, corresponding to the position of the instanton and anti-instanton, respectively, to depend on time. In turn, this will induce time-dependence in the other fields, but let us just consider the gauge-fields for the moment. 

In the gauge $A_y=0$, a pure instanton connection is specified by three potentials $(A_{\th^1}, A_{\th^2}, A_{\th^3})$. Self-duality means that the $3$ components, 
\be
F_{y\th^i} = \p_y A_{\th^i}, 
\ee
determine the entire field strength. This makes a non-abelian system look almost abelian. The instanton configuration depends on $(y-y_1)$. Self-duality also implies a beautiful relation on the stress-energy, 
\be\label{gaugestress}
\tr |F|^2_{mn}= {1 \over 2}g_{mn}  \tr |F|^2,
\ee
which we used in section~\ref{onetransversedimension}. The appropriate metric appearing in~\C{gaugestress}\ is the leading order flat metric~\C{leadingmetric}. 

Let $y_i=y_i(t)$ depend on time.  We still need to solve the Yang-Mills equations of motion, which now include an electric field. We can keep the problem intrinsically $4$-dimensional by taking $A_0=0$. The electric field is then given by,
\be
F_{0\th^i} = {\dot y_1} F_{y\th^i}. 
\ee  
The Yang-Mills equations of motion, $D^\mu F_{\mu\nu}=0$, still take a nice form:
\be\label{modifiedYM}
- \ddot y_1 F_{y\th^i} +  (1- ({\dot y_1})^2) \p_y F_{y\th^i} + D_{\th^j} F_{\th^j \th^i} =0, \qquad D_{\th^j} F_{\th^j y} =0.
\ee
If $\dot y_1=0$, these are the usual $4$-dimensional Yang-Mills equations solved by a self-dual connection. If the acceleration is zero, $\ddot y_1=0$, the resulting equations are a very close cousin to the usual self-duality equations. Constant velocity therefore appears to be a very natural condition.

For a time-dependent configuration with constant velocity for both the instanton and anti-instanton, there is now at least one space-time stress-energy source:
\be
\tr |F|^2_{00} = F_{0\th^i} F_{0{\phantom \th}}^{\phantom{0}\th^i} = (\dot y_1)^2 (F_{y\th^i}(A_1))^2+ (\dot y_2)^2 (F_{y\th^i}(A_2))^2. 
\ee
There could also have been a source $\tr |F|^2_{0\th^i}$, but that off-diagonal stress-energy contribution vanishes for a connection satisfying the modified self-duality constraint associated to~\C{modifiedYM}.
Solving the space-time Einstein equations will now require a more general metric than a maximally symmetric space-time with a single warp factor $w_1$. It is going to be very interesting to explore this direction further though we will restrict to the time-independent ansatz for the remainder of this analysis.

\subsubsection{The $B$-field and dilaton EOMs}

The solution to the $B$-field equation of motion~\C{bfield}\ is that 
\be
H = e^{2\phi} \ast dS = e^{2\phi+8w_1}\hat\ast_4 dS,
\ee
where $S=S(y^m)$ is a scalar field, and $\hat\ast_4$ is with respect to the hatted metric~\C{definehat}. Plugging this into the Bianchi identity~\C{nonsusybianchi}\ gives a Laplace-like equation for $S$:
\be\label{equationforS}
d \left( e^{2\phi+8w_1}\hat\ast_4 dS \right)= - {\a' \over 4}\left\{  \mathrm{tr}\left(F _1\wedge F_1 \right)+ \mathrm{tr}\left(F_2 \wedge F_2 \right)  -  \mathrm{tr}\left(R_+ \wedge R_+ \right)\right\}. 
\ee
In the supersymmetric case, $S = - e^{-2\Phi}$ for an instanton; however, that need not be the case in this non-supersymmetric setting.

There is now a sharper tension between the Bianchi identity and the dilaton equation of motion than was present in the supersymmetric setting:
\be\label{dilagain}
\nabla^2 e^{-2\Phi} = e^{-2\Phi} \left\{ \left|H\right|^2 + {\alpha ' \over 4}\left( \mathrm{tr}|F|^2 - \mathrm{tr}|R_+|^2\right) \right\}.
\ee
While the charge of the instanton and anti-instanton cancel in~\C{equationforS}, they add in the dilaton equation~\C{dilagain}. The contributions to both equations are still related to one another because the instanton and anti-instanton are both BPS configurations. On the other hand, the $|H|^2$ term on the right hand side of~\C{dilagain}\ is integrable unlike the supersymmetric case. If we want net charge $0$, for example, on the right hand side of~\C{dilagain}\ we need a non-minimal curvature $R_+$. There seems to be no real tension between this condition and the topological condition~\C{toponR+}; there are plenty of $SO(4)$ connections with zero instanton charge but non-zero energy. 

The remaining equations of motion are the internal components of the Einstein equations. They can be written out explicitly for the metric~\C{metricansatz}\ but the expressions are not particularly enlightening. We suspect the right way to proceed is to integrate out the compact $T^3$ and study an effective one-dimensional problem involving only the $y$-direction. The reduction to an ODE problem should make the search for explicit solutions more tractable either analytically or numerically.

\subsection{Domain walls between lower-dimensional vacua}\label{lowerdvacua}

We can extend this construction to build domain walls interpolating between the $T^4$ vacua described in section~\ref{quadquint}. 
The construction of the non-trivial quadruples naturally splits the $T^4$ into an $S^1\times T^3$, with a constant holonomy on the $S^1$, whose coordinate we will choose to be $u$, breaking the $Spin(32)$ gauge group to $Spin(16)\times Spin(16)$. The remaining $T^3$ then supports nontrivial triples in each of the $Spin(16)$ factors of opposite Chern-Simons number. Including a non-compact direction, $y$, as before, the gauge configuration interpolates between a nontrivial quadruple at $y=\infty$ and a trivial, flat connection at $y=-\infty$ with gauge group $Spin(16)\times Spin(16)$.

This configuration has no dependence on the $u$ coordinate, and we can choose the gauge field along the $u$-direction, $A_u$, to have no dependence on the non-compact $y$ coordinate along with no dependence on the $T^3$ coordinates. This amounts to choosing the same holonomy along the $S^1$ in either of the asymptotic $|y| = \infty$ vacua. With such a choice, the $\mathbb{R} \times S^1 \times T^3$ field strength becomes effectively $4$-dimensional, having only support along the $\mathbb{R}\times T^3$ directions. In other words, $F_{u \mu} = 0$ for any $\mu$ index.

This $4$-dimensional field strength splits, as in the $E_8 \times E_8$ case, into a sum of self-dual and anti-self-dual pieces, corresponding to an instanton in one $Spin(16)$ factor and an anti-instanton in the other. In this regard, the quadruple domain wall is very similar to that of the triple.

\subsection{Boundaries and curvatures}\label{intervalcompactification}

To get a feel for some of the physics that can emerge from this construction, let us use a much simpler metric ansatz than~\C{metricansatz}. One very much like the supersymmetric case. The simplest solution to the constraint~\C{determinedilaton}\ from the space-time Einstein equations is to set $w_1=0$.  For simplicity, we will also set $w_3=0$ so the metric takes precisely the supersymmetric form:
\be
ds^2  =   \eta_{\m\n} dx^\m dx^\n +e^{2w_2} \left( dy^2 + L^2 \dd_{ij} d\th^i d\th^j \right). \label{almostsusymetric}
\ee
Any curvature response to the $O(\alpha')$ stress-energy from the Yang-Mills fields will be at least order $\alpha'$. The $H$-field is also $O(\alpha')$ so the leading terms in the dilaton equation~\C{dilagain}\ give the relation,
\be
\p^m\p_m \Phi + {\alpha'\over 8} \tr|F|^2 =0, 
\ee 
with any corrections higher order in $\alpha'$. This does not mean the correction terms are unimportant; they can change the total charge and therefore asymptotic behavior of $\Phi$ but as a first approximation, we will neglect them. 

We can also examine the internal Einstein equations~\C{Einstein}\ for the metric~\C{almostsusymetric}\ neglecting the $R_+$ and $H$ contributions, which are again higher order in $\alpha'$. The internal components of the Einstein equations~\C{Einstein}\ then give the relation, 
\be\label{determinew1}
-g_{mn} \left( \p^p\p_p w_2\right) - 2 \p_m\p_n w_2 + 2\p_m\p_n \Phi - {\alpha'\over 8} g_{mn}  \tr|F|^2 =0, 
\ee
where $g_{mn}$ refers to the simple internal metric of~\C{leadingmetric}. These equations are solved by,
\be
w_2=\Phi. 
\ee
This is exactly the same as the supersymmetric case of section~\ref{onetransversedimension}\ except we no longer have the non-integrable $|H|^2$ contribution that caused the problems described in section~\ref{SUGRAanalysis}, because of the better asymptotic behavior permitted for $H$ by the Bianchi identity~\C{nonsusybianchi}.  

The dilaton is sourced the same way by the instanton located at $y_1$, and the anti-instanton located at $y_2$, with asymptotic behavior 
\be
\Phi = -{\alpha' k\over 16}\left( |y-y_1| + |y-y_2| \right)+ \ldots, 
\ee
where $k$ is defined in~\C{instcharge}. The omitted terms are rapidly decaying in $y$. Since $\Delta(CS)>0$ so that $k>0$, the string coupling $g_s= e^\Phi$ goes to $0$ as $|y|\rightarrow\infty$. 
Let us summarize what we have found at leading order in $\alpha'$: in Einstein frame, the asymptotic metric as $|y|\rightarrow\infty$ behaves as follows:
\be
ds^2 = e^{{\alpha' k\over 32}\left( |y-y_1| + |y-y_2| \right)}\eta_{\m\n} dx^\m dx^\n+ e^{-{3 \alpha' k\over 32}\left( |y-y_1| + |y-y_2| \right)}\left( dy^2 + L^2 \dd_{ij} d\th^i d\th^j \right). \label{asymptoticmetric}
\ee
The asymptotic string coupling behaves as follows,
\be
g_s = e^{- {\alpha' k\over 16}\left( |y-y_1| + |y-y_2| \right)},
\ee
with $k>0$. The only potentially worrisome issue with this background is that the string-frame curvature becomes large near the boundaries. We can see this by examining the Ricci scalar using the convenient expression~\C{usefulexpressionsforsimplecase}: 
\be
R \sim e^{-2\Phi}.
\ee
This means higher curvature effects will become important near the boundaries, which is perhaps not surprising. 

What we would like to know immediately is whether the Planck constant obtained by integrating over the internal four directions is finite or infinite. 
The asymptotic region of the integral, which is the only place a divergence could emerge, gives a contribution:
\be
\int dy d\th^1 d\th^2 d\th^3 \sqrt{g} \sim \int dy e^{-{3 \alpha' k\over 32}\left( |y-y_1| + |y-y_2| \right)} < \infty. 
\ee
Therefore the Planck constant is finite and gravity is localized in six dimensions. 
To understand what has happened to the internal space, let us examine a null geodesic along the $y$-direction. Such a geodesic satisfies,
\be
 -e^{\b|y|} dt^2 + e^{-3\b|y|}dy^2=0,
\ee
with $\b = {\alpha' k \over 16}$. Solving this equation with the initial condition $y(t=0)=0$ yields,
\be
2\b |t| = 1 - e^{-2\b |y|}\quad  \Leftrightarrow \quad |y| = -{1 \over 2\b} \log \left( 1- 2\b |t| \right).
\ee
From this behavior, we see that $0 < |t| < {1\over 2\b}$, so there is a boundary at $t = \pm {1 \over 2\b}$. The $y$-direction has effectively compactified to an interval!

There are some rather interesting features of these solutions, which are of the general form~\C{firstform}. Since the $y$-direction compactifies to an interval of finite proper size, we end up with an internal space with boundaries. The space is geodesically incomplete. Boundaries do appear in several places in string theory. At strong coupling, the $E_8\times E_8$ heterotic string is described by heterotic M-theory, which involves the interval $S^1/\Z_2$~\cite{Horava:1995qa}. One set of $E_8$ gauge bosons is supported on each boundary. Similarly type I$^\prime$, which is the T-dual description of type I string theory on a circle, is described by an interval with $O8$-planes supported at the ends of the interval. Lastly, $(0,2)$ chiral gauge theory in two dimensions, which is expected to describe flux vacua of the heterotic string, involves target spaces with boundaries~\cite{Melnikov:2012nm}. 

The structure we see is intriguing and, we suspect, indicative of a general structure in string theory. Here each boundary supports a string vacuum. The interpolating string coupling has an asymptotic linear dilaton behavior in the $y$ coordinate with the string coupling asymptoting to zero as $|y|\rightarrow \infty$:
\be
\Phi \sim - \alpha' |y|.
\ee  
This is very likely not the right coordinate to use to describe the asymptotic dilaton since the metric is still warped in terms of $y$, yet the construction does suggest a holographic description in analogy with little string theory~\cite{Losev:1997hx, Seiberg:1997zk}\ and AdS/CFT~\cite{Maldacena:1997re}\ since the distance, measured with the Einstein frame metric~\C{asymptoticmetric}, between points fixed in $\R^{5,1}$ diverges at the boundaries. 

\subsection{Some additional comments}

Aside from purely field theoretic questions about instantons on $\R\times T^3$, there are many directions for future investigation. Some appear in section~\ref{intro}. We will list a few more here: how do we define string observables in these domain wall space-times? What can be said about holography for these Minkowski space-times? Do brane-anti-brane interactions make these backgrounds time-dependent at higher orders in $\alpha'$?


In addition to the minimal charge instantons connecting fractional $CS$ invariants, we can consider higher charge instantons that differ from the minimal charge configurations by integer jumps of the $CS$ invariant so that
\be
\int dy \int_{T^3} \, \Tr (F_1\wedge F_1) - \int dy \int_{T^3} \, \Tr (F_2\wedge F_2) = k + (8\pi^2 N_R)N, 
\ee
where $|k|<1$ and $N$ is integer. If these additional integer charge instantons can shrink to zero size, this amounts to adding NS5-branes and anti-NS5-branes to the background. Eventually these branes can presumably annihilate leaving the basic minimal charge domain wall structure we have described. However, whether finite size instantons can shrink to zero size requires a detailed study of the moduli space of instantons which connect fractional $CS$ invariants. 

Again a primary reason the walls we have described are so nice is that they involve field theory instantons that can connect string vacua with no need for topology change. In field theory, topology change involves an infinite energy barrier. The same is not true in string theory so we suspect there should exist intrinsically stringy walls that interpolate between topologically distinct vacua. Indeed toroidal vacua with topologically distinct gauge bundles can be related to vacua with topologically trivial gauge bundles via T-duality~\cite{deBoer:2001px}. Can one describe these intrinsically stringy walls?

\section{Domain Walls Between AdS Vacua}\label{AdScase}

We now turn to the construction of domain walls between AdS vacua. We will start by constructing a wall that connects $AdS_3$ vacua. One possible starting point is a system of F1-strings and NS5-branes in $E_8\times E_8$ heterotic string theory. For that setup, we can use the kind of triple domain walls described in section~\ref{Minkdomain}. However, the holographic dual for this NS system is not as simple as in the case of D-branes. For the purpose of understanding holography for accelerating space-times, it is simpler to consider a D1-D5 system in type I string theory with the D5-branes wrapping $T^4$. In either case, we will have to study the supergravity solution for the 1-brane/5-brane system in either heterotic or type I string theory. 

\subsection{F1-NS5 system in heterotic string theory}  \label{f1ns5vacua}

Let us start by describing the vacua of interest to us. We can study the supergravity solution for the vacuum configuration in either the heterotic frame or the type I frame. The $Spin(32)/\Z_2$ heterotic string and the type I string are S-dual, while the $E_8\times E_8$ heterotic theory is distinct. Nevertheless, for the purpose of solving the equations of motions, the gravity sector for each theory is identical; they only differ in their gauge-field content which will be visible at $O(\alpha')$. 

The parameter map relating the ten-dimensional type I string dilaton $\Phi_I$ to the heterotic dilaton $\Phi$, the type I string metric to the heterotic metric, and relating the heterotic NS $H$-field to the type I RR $F_3$-field appeared earlier in~\C{dualitymap}. For convenience, we reproduce it here:
\be
\Phi_I = - \Phi, \qquad ds^2_{I} = e^{-\Phi} ds^2_{het}, \qquad F_3=H. 
\ee
Since we already have the equations of motion expressed in the heterotic frame~\C{dilatoneom}-\C{gaugeeom}, let us continue our discussion in heterotic variables. 

The F1-NS5 system of the heterotic string requires a mild generalization of the usual supergravity solution describing the type II F1-NS5 system. We will not take a decoupling limit initially. We want to solve the string equations of motion for $n_1$ F1-branes and $n_5$ NS5-branes. The type II solution takes a very nice  form in string frame with metric (found, for example, in~\cite{Mathur:2005zp}),
\be\label{ns15metric}
ds^2_{II} = {1\over f_1}\left( -dx_0^2 + dx_1^2 \right) + f_5\left( dr^2 + r^2d\Omega_3^2 \right) + ds^2_{T^4},
\ee
with 
\be
f_1 = 1 + {Q_1 \over r^2}, \qquad  f_5 = 1 + {Q_5 \over r^2}. 
\ee
The charges $Q_1$ and $Q_5$ are proportional to $n_1$ and $n_5$. The string dilaton is determined in terms of the $f_i$,
\be\label{defstringcoupling}
e^{2\Phi} =g_s^2  {f_5\over f_1},
\ee
with $g_s$ the asymptotic value of the string coupling. There is an $H$-flux that takes the form,
\be
H = 4\pi^2Q_5 \epsilon_3 +{4\pi^2 Q_1 \over g_s^2} e^{2\Phi} \ast_6\epsilon_3,
\ee
where $\e_3$ is the volume form for the unit three sphere normalized so that $\int_{\Omega_3}\e_3 = 1$, and $\ast_6$ is the Hodge star operation for the $6$ directions transverse to the $T^4$. Written in the orthonormal basis of Appendix~\ref{curvatures},
\be\label{defH}
H = {2Q_5 \over r^3 f_5^{3/2} } e^{\theta} \wedge e^{\phi} \wedge e^{\psi} - {2Q_1e^{2\Phi} \over g_s^2 r^3 f_5^{3/2}}e^0\wedge e^1 \wedge e^r,
\ee
where we have used:
\be
\e_3 = { e^{\th} \wedge e^{\phi} \wedge e^{\psi} \over 2\pi^2 r^3 f_5^{3/2}}.
\ee
We will be interested in taking the decoupling limit where we drop the constant in $f_1$ and $f_5$,
\be\label{decoupling}
f_1\rightarrow {Q_1\over r^2}, \qquad f_5\rightarrow {Q_5\over r^2}, 
\ee
and the space-time becomes $AdS_3$. 

Since the difference between heterotic and type IIB string theory involves terms of $O(\alpha')$, this structure determines the heterotic solution at leading order. Of course in the heterotic string, there are more ways of generating $n_5$ using fat gauge-field instantons in addition to branes, but this solution still describes the leading order supergravity solution. For large $(n_1, n_5)$, the background has small curvatures and we can trust an $\alpha'$ expansion. At this order, the gravity solutions corresponding to the different heterotic or type I vacua are distinguished only by the choice of $Spin(32)/\Z_2$ or $E_8\times E_8$ flat connection on $T^4$: either a quadruple configuration or a triple configuration, both with zero field strength.

\subsection{Choices in building domain walls}

We want to  mimic the Minkowski space construction of section~\ref{Minkdomain}. Before worrying about an interpolating solution, we note that the Bianchi identity now has a non-trivial gravitational contribution that, in principle, cannot be neglected:
\be\label{fullbianchi}
d H =  {\a' \over 4}\left\{  \mathrm{tr}\left(R(\Omega_+) \wedge R(\Omega_+) \right)- \mathrm{tr}\left(F \wedge F \right) \right\}.
\ee
The connection, $\Omega_+$, used to evaluate the metric curvature was defined earlier in~\C{conn}; it is a combination of the usual spin connection $\Omega$ and $H$:
\be\label{repeattorsionconnection}
\Omega_+ = \Omega + {1\over 2} H. 
\ee
This is an added complication in the heterotic system not present in the type II solution. A closely related complication appears in the dilaton and Einstein equations,~\C{dilatoneom}\ and~\C{Einstein}, which now involve an $|R_+|^2$ source term that is now non-trivial because of the background metric~\C{ns15metric}.  

Fortunately, there is a very nice way to reduce the analysis to a situation almost as simple as the Minkowski case we studied earlier. The ideal case would be a construction analogous to the symmetric $5$-brane solution, which makes the right hand side of~\C{fullbianchi}\ vanish by identifying the gauge connection with the spin connection. This choice is not possible in a fully Lorentzian background like the case considered here. However, the situation is actually better! 

It is better in two ways. In Appendix~\ref{curvatures}, we have computed the torsionful spin connection~\C{repeattorsionconnection}. This connection depends on the radial functions~\C{radialfunctions}. However, it is easy to see that the connection vanishes for large $Q_1$ and $Q_5$. So in a large charge limit, the gravitational sources will be subleading when compared with the instanton sources we use to build the domain wall, which are $O(1)$ in the charge expansion. 

Even if we choose not to take a large charge limit, we can still construct an analogue of the heterotic standard embedding. The expression for the gravitational contribution to~\C{fullbianchi}\ computed in Appendix~\ref{curvatures}\ is given below:
\be\label{repeatgravcontribution}
 \tr(R_+ \wedge R_+) =  \frac{16 Q_5^2  \sin \psi
   \left(f_5(\sin \theta \sin \psi - \cos \psi) + 3 \sin\theta\sin \psi\right)}{r^5 f_5^4}dr \wedge d\th \wedge d \phi \wedge d\psi.
\ee
 This expression is completely independent of $Q_1$. It depends only on $Q_5$ and the $(r,\th,\phi,\psi)$ directions. The Lorentzian features of the this background all depend on $Q_1$ and therefore play no role. At worst, we can choose an $SO(4)$ connection to cancel the gravitational contribution to~\C{fullbianchi}. For example, we can take an $SU(2)$ factor from each $E_8$ of $E_8\times E_8$, or $SO(4)$ directly from $SO(32)$. In the former case, we have an unbroken $E_7\times E_7$ gauge group while in the latter, an unbroken $SO(28)$ gauge group. 

However, this is the worst case. Since the background metric~\C{ns15metric}\ in the $(r,\th,\phi,\psi)$ directions is simply the conventional NS5-brane solution, with no $Q_1$-dependence, we should be able to construct a left-right symmetric world-sheet CFT for those directions using only an $SU(2)$ connection rather than $SO(4)$. For the $SO(32)$ string, this improvement will not matter for our subsequent discussion because any $SO(N\geq 7)$ unbroken gauge group has a unique non-trivial triple vacuum. We can therefore easily construct a quadruple vacuum for the $SO(32)$ string along the lines described in section~\ref{quadquint}.  

For the case of the $E_8\times E_8$ string, we have an unbroken gauge group which is at worse $E_7\times E_7$ and possibly as large as $E_7\times E_8$. The moduli space for the group $E_7$ on $T^3$ consists of $6$ disconnected components labeled  by a $CS$ invariant with possible values: $\left(0, {1\over 4}, {3\over 4}, {1\over 3}, {2\over 3}, {1\over 2} \right).$ Once again, we simply embed equal and opposite $CS$ invariants in each group factor to form a string vacuum on $T^3$; we can also repeat the construction described in section~\ref{quadquint}\ to construct quadruples, and even quintuples if desired. 

The upshot of using this partial analogue of the standard embedding is that we can simply forget about the gravitational contribution to~\C{fullbianchi}\ and the gravitational source $|R_+|^2$, as long as we use the residual unbroken gauge group to build vacua and interpolating instantons.  This puts us in a situation essentially as good as the Minkowski case. To proceed, we need to choose a direction, whose coordinate we called $y$ in our prior discussion, along which to build the interpolating instanton. Unlike the Minkowski case, there are several choices in this case:
\begin{itemize} 
\item 
We can replace $T^4$ with coordinates $(\th^1, \th^2, \th^3, \th^4)$ by $T^3\times \R$ with coordinates denoted $(\th^1, \th^2, \th^3, x_4)$.  We are then free to interpolate along the $x_4$ direction. Clearly, we can only interpolate between triple vacua so this choice requires the $E_8\times E_8$ heterotic theory. This choice preserves the $Spin(1,1)$ Lorentz symmetry acting in the $(x_0, x_1)$ directions, along with the $Spin(4)_R$ symmetry acting in the $(r, \th,\phi,\psi)$ directions. 

\item 
We can interpolate along $x_1$. This is the closest analogue to our Minkowski domain wall discussion, and the easiest case to interpret holographically. This choice breaks the $Spin(1,1)$ Lorentz symmetry, but preserves the $Spin(4)_R$ symmetry. 

\item
We can interpolate along the $r$-direction. This choice preserves both  the $Spin(1,1)$ Lorentz and $Spin(4)_R$ symmetries. 

\item
We can choose a spatial direction transverse to the brane system. This choice breaks $Spin(4)_R$ to $Spin(3)_R$, but preserves the $Spin(1,1)$ Lorentz symmetry.

\item
We can interpolate along $x_0$. This choice again breaks the $Spin(1,1)$ Lorentz symmetry, but preserves the $Spin(4)_R$ symmetry. It is a domain wall in real time rather than a spatial direction. 
\end{itemize}
If we choose to preserve the $Spin(4)_R$ rotational symmetry that acts in the directions transverse to the NS5-branes, we could also consider a linear combination of the $r$ and $x_1$ directions, but it is simpler to first consider these two choices separately. 
The last case of interpolating in time is an option we could have also studied in the Minkowski case. The physical nature of the setup is quite different from a conventional domain wall since the gauge field configuration is Lorentzian rather than Euclidean; we will not examine that possibility further here.

\subsection{Interpolating along $x_4$}

This turns out to be the easiest case to analyze because we can use a great deal of our prior analysis in the Minkowski case. The torus metric appearing in~\C{ns15metric}\ has no knowledge of the branes. Let us replace $T^4$ by $T^3\times \R$ with coordinates $(\th^1, \th^2, \th^3, x_4)$. We will take a nice simple form for this metric, 
\be
ds^2_{T^3\times \R} = ds^2_{T^3} + dx_4^2. 
\ee
We must have collection of potentials, $A_{\th^i}(x_4,\th^i)$ which interpolate from one vacuum configuration at $x_4=-\infty$ to another vacuum at $x_4=+\infty$. We are identifying the $y$-direction of our prior discussion with the $x_4$ direction.  For this interpolation choice, we can simply use the instanton/anti-instanton configurations we described in the Minkowski case to solve the gauge field equation of motion. This is a very nice simplification. It means we can basically add the domain wall solution described in section~\ref{intervalcompactification}\ to the supergravity solution for $Q_1$ fundamental strings smeared along the $x_4$ and torus directions, and $Q_5$ NS5-branes wrapping these directions. The result is a remarkably simple and straightforward combination of the Minkowski domain wall, and the standard F1-NS5 solution.   

Let us denote the fields associated with the F1-NS5 solution with the superscript ``NS''; similarly, ``DW'' will denote Minkowski domain wall fields. The combined solution is just the sum of these components:
\be
F = F^{\rm NS} + F^{\rm DW}, \, \ H = H^{\rm NS} + H^{\rm DW}, \, \ \Phi = \Phi^{\rm NS}  + \Phi^{\rm DW},
\ee
with metric
\be\label{transversebrane}
ds^2 = {1\over f_1}\left( -dx_0^2 + dx_1^2 \right) + f_5\left( dr^2 + r^2d\Omega_3^2 \right) + e^{2\Phi^{\rm DW}}\left( ds^2_{T^3} + dx_4^2 \right).
\ee
To show that this is a solution, we note that $H^{\rm NS}$ and $F^{\rm NS}$ are supported only in directions transverse to their $DW$ counterparts, and the two contributions to $\Phi$ have dependence on non-overlapping sets of coordinates. Start with the $B$-field equation of motion:
\bea
d\left( e^{-2\Phi} \ast H \right) &=& e^{2\Phi^{\rm DW}}d \left ( e^{-2\Phi^{\rm NS}} \ast_{\rm NS} H^{\rm NS} \right) +e^{-2\Phi^{\rm NS}}{f_5^2 \over f_1} d\left( e^{-2\Phi^{\rm DW}}\ast_{\rm DW} H^{\rm DW}\right), \cr &=& 0.
\eea
We use $\ast_{\rm DW/NS}$ to refer to the Hodge star operation on the purely Minkowski domain wall metric and the standard F1-NS5 metric without a domain wall solution, respectively. This equation of motion is satisfied using the respective equations of motion for the NS and DW $B$-fields. For this factorization to be true, it is crucial that:
\be
d\Phi^{\rm DW} \wedge \ast H^{\rm NS} = 0,
\ee
with a similar relation where DW and NS are reversed. For similar reasons, the gauge equation of motion splits as follows:
\be
e^{4\Phi^{\rm DW}} \left( F^{\rm NS} \text{ E.O.M.} \right) + {f_5^2 \over f_1} \left( F^{\rm DW} \text{ E.O.M.} \right) = 0.
\ee
Furthermore,
\bea
& |H|^2 = |H^{\rm NS}|^2 + |H^{\rm DW}|^2, \qquad  \tr|F|^2 = \tr|F^{\rm NS}|^2 + \tr|F^{\rm DW}|^2, & \cr & R = R^{\rm NS} + R^{\rm DW}, & \non
\eea
so the dilaton and Einstein equations are also satisfied if the NS and DW equations are separately satisfied.

Note that the harmonic function $f_1$ appearing in~\C{transversebrane}\ should be the one appropriate for the actual volume of $T^3\times \R$, taking into account the domain wall back reaction. When the domain wall leads to a finite volume space, along the lines discussed in section~\ref{intervalcompactification}, then the harmonic function for  a finite volume $T^4$ should be used. This is interesting and a little surprising. Suppose we had imagined building this background sequentially. Start first with the F1-NS5-brane solution compactified on a $T^3\times \R$ transverse to the fundamental strings but parallel to the NS5-branes. The metric takes the form,
\be
ds^2 = {1 \over \widetilde{f}_1}( -dx_0^2 + dx_1^2) + f_5( dr^2 + r^2 d\Omega_3^2) + dx_4^2 + ds_{T^3}^2,
\ee
where 
\be
\widetilde{f}_1 = 1 + {Q_1 \over (r^2 + x_4^2)^{3/2}}
\ee
is harmonic in the $5$ non-compact directions transverse to the fundamental strings. This would not give $AdS_3$ in the near horizon limit. Next insert the  instanton/anti-instanton configuration along $T^3\times \R$. This procedure must reproduce the metric found in~\C{transversebrane}, changing $\widetilde{f}_1$ to $f_1$. Interestingly, this is an $O(1)$ response to the gauge-field configuration, whose effects are naively $O(\alpha')$. The reason this is happening is that the domain wall is basically a collection of particles in one dimension. The gravitational back reaction produced by those particles is large, regardless of any $\alpha'$ suppression. It is sufficiently large that it renders the volume of the line finite as we saw in section~\ref{intervalcompactification}.

\subsection{The gauge field equation of motion}

Interpolating along any other direction already introduces new issues with solving the gauge-field equation of motion, which are quite fascinating. Here we will describe  the new questions that arise. 
We again assume a nice simple form for the $T^4$ metric appearing in~\C{ns15metric}, 
\be
ds^2_{T^4} = ds^2_{T^3} + (d\th^4)^2. 
\ee
Following the discussion in section~\ref{lowerdvacua}, the field strength for the gauge-fields has no support in the $\th^4$ direction. As in section~\ref{gaugeeommink}, we will expand all fields, like the dilaton, in a perturbative expansion in $\alpha'$:
$$
\Phi = \Phi^{(0)} + \alpha' \Phi^{(1)} + \ldots. 
$$
Since the background fields $\Phi^{(0)}$ and $H^{(0)}$ are now non-trivial, we need to take a step back and reconsider the interpolating gauge-field configuration that forms the basis for our domain wall backgrounds. 
The heterotic gauge field equation of motion~\C{gaugeeom}\ is now a fully $10$-dimensional equation, which requires a connection that solves
\be\label{fullproblem}
D \left( e^{-2\Phi^{(0)}} \ast F\right) +e^{-2\Phi^{(0)}} F\wedge \ast H^{(0)}_{Q_1}=0, 
\ee
with the full $10$-dimensional Hodge star. The $F\wedge\ast H^{(0)}$ final term of~\C{gaugeeom}\ is only non-vanishing for the electric term in $H^{(0)}$, proportional to $Q_1$, which we have denoted $H^{(0)}_{Q_1}$.  We stress that the terms in $H$ and $\Phi$ of $O(1)$ in the $\alpha'$ expansion were not present in the Minkowski case. 

Let us write out equation~\C{fullproblem}\ explicitly using the conventions of Appendix~\ref{curvatures}. Nothing depends on $(\th, \phi, \psi, \th^4)$ so we will drop the volume form in those directions,  
\bea\label{fullequation}
&& D \left\{  (r^3 f_5 f_1) F_{x_1\th^i} dr dx_0 ({1\over 2} \e_{ijk} d\th^j d\th^k)   + (r^3 f_5) F_{\th^j \th^k} dr dx_0 dx_1 ({1\over 2}\e_{ijk} d\th^i) \right.  \cr 
&& \left.  + (r^3) F_{r \th^i} dx_0 dx_1 ({1\over 2}\e_{ijk} d\th^j d\th^k) - (r^3 f_1) F_{r x_1} dx_0 d\th^1 d\th^2 d\th^3 - (r^3 f_1) F_{r x_0 } dx_1 d\th^1 d\th^2 d\th^3  \right.  \cr 
&& \left. - (r^3 f_1^2 f_5) F_{x_0 x_1} dr d\th^1 d\th^2 d\th^3 + (r^3f_5f_1) F_{x_0 \th^i} dr dx_1 ({1\over 2}\e_{ijk} d\th^j d\th^k) \right\}  \cr 
&& + \left\{ F_{rx_1}dr dx_1 + F_{rx_0}dr dx_0 + F_{x_0 x_1}dx_0 dx_1 \right\} (2Q_1 d\th^1 d\th^2 d\th^3)=0.  
\eea
The covariant derivative $D$ only acts in the $(r, x_0, x_1, \th^1, \th^2, \th^3)$ directions. This is a quite complicated looking collection of PDEs, which we will attempt to unentangle. 

\subsubsection{Interpolating along $x_1$}

If we choose to interpolate along the $x_1$ direction then the field strength has support in the $(x_1, \th^1, \th^2, \th^3)$ directions, where $(\th^1, \th^2, \th^3)$ are coordinates for $T^3$. However, this is not sufficient to construct a solution. The metric for the $x_1$ direction depends on $r$ so a purely $4$-dimensional gauge field configuration will not suffice to solve~\C{fullequation}, as we will see below. The asymmetric warping between the $T^3$ and the $x_1$ direction in the metric~\C{ns15metric}\ is the basic source of all the new issues. 

At least initially, we might have expected a $5$-dimensional holographic extension in the $(r, x_1, \th^1, \th^2, \th^3)$ directions of our $4$-dimensional instanton/anti-instanton configuration living in the $(x_1, \th^1, \th^2, \th^3)$ directions. However, the  electric $H^{(0)}_{Q_1}$ term mixes the time direction into the equations. So initially, we must retain the possibility of a $6$-dimensional gauge-field configuration in the  $(r, x_0, x_1, \th^1, \th^2, \th^3)$ directions.

What we must have is a collection of potentials $A_{\th^i}(r,x_1,\th)$ which interpolate, at least at the boundary located at radial infinity, from one vacuum configuration at $x_1=-\infty$ to another vacuum at $x_1=+\infty$. The complication that we face in solving~\C{fullequation}\ is that the metric is warped in the $x_1$ direction, but not the $\th^i$ directions. This is what forces us to have radial dependence. At a minimum, we have non-vanishing field strengths $(F_{\th^i\th^j}, F_{r\th^i}, F_{x_1\th^i})$. 

In general, there is no reason to expect a particularly simple solution to~\C{fullequation}. For example, if we need to turn on an $A_{x_1}$ potential, which leads to an $F_{rx_1}$ field strength, it looks likely that we will also need an $A_{x_0}$ potential, and most terms in~\C{fullequation}\ will contribute. The problem becomes Lorentzian rather than being purely Euclidean because of the $Q_1$ charge. A similar comment applies if turn on an $A_r$ potential. We can choose the gauge, 
\be\label{axialgauge}
A_{x_1}=0, 
\ee
or $A_r=0$ to kill one of these two possibilities, but not both together. At the boundary $r=\infty$, we do expect to be able to impose both conditions: $A_{x_1}=A_r=0$. 

However, it is worth at least exploring the possibility of a simple solution. In addition to the gauge choice~\C{axialgauge}, if we can also maintain $A_{r}=0$ so no $F_{rx_1}$ field strength is produced then the first $3$ terms of~\C{fullequation}\ form the following closed system of equations:
\bea \label{firstYM}
& D_{\th^i} F_{x_1\th^i} = D_{\th^i} F_{r\th^i} =0, &\\ & (r^3 f_5 f_1) \p_{x_1} F_{x_1\th^i} + \p_r \left(r^3 F_{r\th^i} \right) + (r^3 f_5) D_{\th^j} F_{\th^j \th^i} =0. \label{secondYM} &
\eea 
It is probably asking too much to be able to find solutions with both $A_{x_1}=0$ and $A_r=0$, but we will see that any $A_r$ needed for an exact solution is subleading in a formal large $Q_1,Q_5$ expansion. 

We will be interested in solutions that survive the decoupling limit~\C{decoupling}, where we drop the constant in $f_1$ and $f_5$. In this limit, equation~\C{secondYM}\ becomes:
\be
 \p_{x_1} F_{x_1\th^i} +  {r^2\over Q_1} D_{\th^j} F_{\th^j \th^i} + {r\over Q_1Q_5}\p_r \left(r^3 F_{r\th^i} \right) =0.\label{decoupledYM}
\ee
We $(Q_1, Q_5)$ both large while ${Q_5\over Q_1}$ remains small so that $g_s$, given in~\C{defstringcoupling}, is small. In this large charge limit for which an $\alpha'$ expansion makes sense, we can try balancing the first two terms of~\C{decoupledYM}. In this approach, we view the last term as formally smaller because of its additional $Q_5$ suppression.  

It is useful to recall how the usual instanton solution solves the equations of motion. A solution to the self-duality equations with respect to the unwarped metric,
\be
ds^2 = dy^2 + (d\th^1)^2+ (d\th^2)^2+ (d\th^3)^2,
\ee
in the gauge gauge $A_y=0$ requires a collection of potentials, $A_{\th^i}$, with field strengths satisfying:
\be\label{bdrydata}
F_{y\th^i} ={1\over 2}  \e_{ijk} F_{\th^j\th^k}. 
\ee
The $d\th^1 d\th^2 d\th^3$ term of the Bianchi identity $DF=0$ then implies that $D_{\th^i} F_{y\th^i} =0$.  Similarly, the $dy d\th^i d\th^j$ terms of Bianchi are equivalent to the remaining equations of motion. The integral $\int F\wedge F$ captures the topology of the gauge-field configuration. Using this solution, we can solve~\C{decoupledYM}\ to leading order in a large charge expansion. Let us view $y=y(x_1, r)$ so that
\be
F_{x_1\th^i} = {\p y \over \p x_1} F_{y\th^i}(y, \th) =  {\p y \over \p x_1} {1\over 2}  \e_{ijk} F_{\th^j\th^k} . 
\ee
The first equation of~\C{firstYM}\ is then automatically satisfied. Writing out~\C{decoupledYM}\ gives, 
\be\label{YMsimpler}
\left[ {\p^2 y\over \p x_1^2} + \left({\p y \over \p x_1}\right)^2 \p_y \right] F_{y\th^i}   - {r^2\over Q_1} \e_{ijk} D_{\th^j} F_{y\th^k} + {r\over Q_1Q_5}\p_r \left(r^3 F_{r\th^i} \right) =0.
\ee
At leading order in the large charge expansion, we choose
\be\label{redefiney}
y= {r\over \sqrt{Q_1}} x_1,
\ee
so that the first two terms of~\C{YMsimpler}\ cancel because of the Bianchi identity.  The price we pay for the choice~\C{redefiney}\ is the generation of a field strength in the $r$ direction, 
\be
F_{r\th^i} =   {\p y \over \p r} F_{y\th^i}(y, \th) = {x_1\over \sqrt{Q_1}} F_{y\th^i}(y, \th). 
\ee
This field strength solves the second equation of~\C{firstYM}. In this attempt at finding a simple leading order solution, the way $y$ depends on $(r,x_1)$ in~\C{redefiney}\ looks potentially problematic at $r=0$. This problem might be resolved at higher orders in the large charge expansion or by a different ansatz for the form of the solution, but really a more powerful approach is needed for establishing the existence of interpolating solutions for~\C{fullequation}\ beyond a perturbative analysis.\footnote{Minimizing an appropriate energy functional with critical points solving~\C{fullequation}\ would be a natural way to proceed. This is a subtle question because of the potentially Lorentzian nature of the equation!}


\subsubsection{Interpolating along $r$}

Let us briefly consider the third choice where we interpolate along $r$. 
At leading order, we will assume the gauge potentials are time-independent. Therefore, the field strength for the instanton/anti-instanton configuration is supported in the $(r, \th^1, \th^2, \th^3)$ directions, where $(\th^1, \th^2, \th^3)$ are again coordinates for $T^3$. The first terms in the gauge field equation of motion~\C{gaugeeom}\ now requires a connection that solves,
\be\label{effectiveproblem}
D \left( e^{-2\Phi^{(0)}} \ast F\right) =0, 
\ee
with the full $10$-dimensional Hodge star. The $F\wedge\ast H^{(0)}$ final term of~\C{gaugeeom}\ vanishes because $\ast H^{(0)}$ always involves the volume form of $T^4$. 

The equation~\C{effectiveproblem}\ does not reduce to any simple covariant $4$-dimensional problem. We can reduce the equation to a non-covariant $4$-dimensional system as follows: choose a new coordinate, $y$, so that the radial metric is canonical: 
\be
dy^2 = f_5 dr^2. 
\ee 
Explicitly, 
\be
y(r) = \sqrt{Q_5+r^2} + \sqrt{Q_5} \log({r\over Q_5}) - \sqrt{Q_5}\log\left(1+ \sqrt{1+ {r^2\over Q_5}}\right). 
\ee
More useful are the two limiting behaviors, 
\be
r\rightarrow 0, \quad y\sim \sqrt{Q_5} \left( 1 + \log({r\over 2Q_5})\right), \qquad\qquad r\rightarrow \infty, \quad y\sim r - {\sqrt{Q_5}\over 2} \log(Q_5). 
\ee
The coordinate $y$ ranges from $(-\infty, \infty)$, which corresponds to $r$ ranging from $(0,\infty)$. Let $\hat\ast$ denote a $4$-dimensional Hodge star with respect to the metric, 
\be
{\widehat{ds}}^2 = dy^2 + (d\th^1)^2+ (d\th^2)^2+ (d\th^3)^2. 
\ee
We can rewrite~\C{effectiveproblem}\ as a $4$-dimensional problem, 
\be\label{rewritten}
D\, \hat{\ast} \, ( r^3 \sqrt{f_5}F)=0. 
\ee
This equation does not correspond to a Yang-Mills field coupled to a background metric. For large and small $r$, we note that
\be
r\rightarrow \infty, \quad r^3 \sqrt{f_5}\sim r^3 \sim y^3, \qquad\qquad r\rightarrow 0, \quad r^3 \sqrt{f_5}\sim r^2 \sim e^{{2y \over \sqrt{Q_5}}}. 
\ee
The extra radial factor in~\C{rewritten}\ does not require the field strength to decay more rapidly than the exponential decay we might have expected at large $|y|$; in fact, the opposite is true. Finite energy now requires,
\be
\int dr (r^3 f_5) |F|^2 < \infty,   
\ee 
which again looks like a reasonable requirement. Actually in this case,  it is not completely clear we need to insist on finite energy as long as the field strength is not too badly behaved as $r\rightarrow 0$. If one is willing to give up the Bianchi identity, it is actually not hard to relate solutions of~\C{rewritten}\ to rescaled conventional instanton and anti-instanton field strengths. The price one pays for relaxing the Bianchi identity is the introduction of a magnetic source. If we insist on satisfying the Bianchi identity then the existence question is quite fascinating but beyond the scope of this work. 

\subsection{Comments on holography}

One of our main goals in this work was to identify a string framework which might provide a definition of quantum gravity in accelerating space-times, or at least in new space-times beyond the handful of examples for which holography is presently understood. While there are many existence questions yet to be addressed, the framework we have described does suggest the general form of a holographic dual description.  

Let us start by discussing the type I and heterotic vacuum configurations described in section~\ref{f1ns5vacua}. Decoupling limits in type I and heterotic string theory are more subtle than their type II counterparts; they have been explored recently  in~\cite{Sethi:2013hra}. For example, the theory of type I D1-strings in a decoupling limit that keeps the Yang-Mills coupling on the branes finite while $g_s\rightarrow 0, \alpha' \rightarrow 0$ is expected to give a compactified $2+1$-dimensional theory rather than the $1+1$-dimensional gauge theory we might have expected based on type II intuition. 

For the type I D1-D5 system, we want to take the decoupling limit that results in $AdS_3$~\cite{Maldacena:1997re}. It is difficult to find a linear theory governing the dynamics of the D1-D5 system because the D5-branes are wrapped on a compact $T^4$. This is already true for the type IIB theory, without any of the additional issues that type I brings. Had the D5-branes been wrapped on $\R^3\times S^1$ or $\R^2\times T^2$ then we could have provided a linear description in terms of an impurity or defect gauge theory; the bulk theory would be $2+1$ or $3+1$-dimensional, respectively, with $1+1$-dimensional defects~\cite{Sethi:1997zza, Kapustin:1998pb}.   

Instead, we will follow the conventional practice of discussing the non-linear theory. For the type II D1-D5 system, it is a $(4,4)$ sigma model with a target space that is a deformation of the orbifold
\be
(T^4)^{n_1n_5}/S^{n_1n_5}, 
\ee
where $S^{n_1n_5}$ denotes the permutation group. 

For the type I theory, we expect a similar picture in which the non-linear theory describes the dynamics of $n_1$ instantons of $Sp(n_5)$ gauge theory\footnote{We are following the convention where $Sp(n)$ refers to the group of rank $n$. }.  We note that the dimension of the moduli space, $\M_{n_1}(G)$, of $n_1$ instantons of a group $G$ on $\R^4$ is 
\be {\rm dim}\, \M_{n_1}(G) = 4n_1h(G), \ee 
where $h(G)$ is the dual Coxeter number of $G$.  For $G=Sp(n_5)$, $h(G)=n_5+1$. 

The $1-1$ strings give an $O(n_1)$ vector multiplet and $8$ real scalars transforming in the symmetric representation of  $O(n_1)$, together with the right-moving fermions required by supersymmetry.  The theory on the D1-strings is an $O(n_1)$ gauge theory with $(0,4)$ supersymmetry. 
The $1-9$ strings give left-moving chiral fermions, $\g$, transforming in the $(n_1, 1, 32)$ of $O(n_1)\times Sp(n_5) \times SO(32)$. These are the only fields that detect the $SO(32)$ global symmetry, and therefore the only fields sensitive to a background type I $SO(32)$ connection.   

So far, we have described the field content of the $(0,8)$ theory supported on type I D1-strings. In addition, we have a hypermultiplet from the $1-5$ strings transforming in the $(n_1, 2n_5, 1)$. The expectation values for this field form the Higgs branch describing the phase where the D1-strings dissolve in the D5-branes. Extending this picture to the compact case, we expect the Higgs branch metric, $\M$, to be a deformation of 
\be
(T^4)^{n_1(n_5+1)}/S^{n_1(n_5+1)}. 
\ee
The full orbifold group for the theory also includes an action on the $\g$ fermions. See~\cite{Gava:1998sv, Banks:1997it, Rey:1997hj, Lowe:1997sx}\ for a discussion in the case of just D1-strings. 
It is now very reasonable that the non-trivial vacuum involving the $SO(32)$ quadruple on $T^4$ is realized by a non-linear sigma model on $\M$ with left-moving fermions coupled to the non-trivial $SO(32)$ gauge connection for the quadruple.   

Finally, we can describe the holographic dual for the interpolating domain wall configurations.  
\begin{itemize} 
\item Consider the case where we replace $T^4$ by $T^3\times \R$ with coordinates  $(\th^1, \th^2, \th^3, x_4)$, and interpolate along the $x_4$ direction.  The $Spin(1,1)$ Lorentz symmetry is preserved so we expect a $1+1$-dimensional sigma model with a target space that reflects the closed string physics we described in section~\ref{Minkdomain}. For example, the $\R$ direction can compactify to an interval as discussed in section~\ref{intervalcompactification}. For backgrounds where the $\R$ direction does not compactify, we do not expect the space-time solution to asymptote to AdS with a CFT holographic description.

\item 
We can interpolate along $x_1$. In this case, we are allowing the parameters of the $SO(32)$ gauge bundle, described by the couplings of the left-moving fermions $\g$, to vary with the spatial $x_1$ coordinate. This choice breaks the $Spin(1,1)$ Lorentz symmetry. The bubble of $1+1$-dimensional domain wall space-time sitting inside $AdS_3$ should be described by this non-Lorentz invariant field theory. In general, the domain wall configuration is also time-dependent. The holographic dual is then a sigma model with parameters that vary with $(x_0, x_1)$.   

This kind of structure has appeared in past holographic descriptions of cosmological space-times; usually in cases where the cosmology involves a dependence on null-time.  The first example of this sort was of AdS/CFT type where both the boundary theory, which involved space-time non-commutativity, and the gravity dual  had dependence on null time~\cite{Hashimoto:2002nr, Das:2006pw}. Later examples are of the matrix big bang flavor~\cite{Craps:2005wd, Craps:2006yb, Blau:2008bp, Das:2007dw,  Das:2008zzf, Craps:2011sp}\ for which the space-time cosmology involves null-time-dependence, but the matrix model involves a field theory with time-dependent parameters.  Field theories with space-time-dependent parameters exhibit interesting phenomena not seen in their Lorentz invariant counterparts; see, for example~\cite{Craps:2006xq, Lin:2006sx, Martinec:2006ak, Dong:2012ua}. 

\item
The last case we will discuss is interpolation along the $r$-direction. This choice preserves  the $Spin(1,1)$ Lorentz symmetry. Usually, we expect motion in the $r$ direction to correspond to an RG flow of some sort. We do not know if static interpolating Yang-Mills configurations exist in this case; if they do and the full gravity solution is static, the holographic interpretation looks quite mysterious. If time-dependent solutions exist, on the other hand, one could imagine perturbing the UV (large $r$) $1+1$-dimensional theory with a time-dependent operator to generate a flow between distinct Yang-Mills vacua on $T^4$.    
\end{itemize}

\section*{Acknowledgements}

It is our pleasure to thank Pierre van Baal, Benoit Charbonneau, Andreas Karch, David Kutasov, Andrei Parnachev, Mark Stern and Dan Waldram for helpful discussions.  Early versions of this construction were presented by S.~S. at the ``de Sitter Days" workshop, Fermilab, 2003 and the Solvay/APC/PI ``Workshop on Cosmological Frontiers," Brussels, 2009. S.~S. would like to thank the organizers of those workshops.  T.~M. and S.~S. are supported in part by
NSF Grant No.~PHY-1316960.  

\newpage
\appendix

\section{Some Useful Relations} \label{usefulformulae}
\subsection{Conformal transformations}
For a $D$-dimensional metric of the form,
\be\label{conformalmetric}
g_{MN} = e^{2\omega}\hat g_{MN},
\ee
 the curvatures can expressed in terms of curvatures of $\hg$ along with derivatives of the conformal factor:
\bea
R_{MNP}{}^Q &=& \hat{R}_{MNP}{}^Q - 2\hn_{[M}^{}C_{N]P}^Q + 2C^R_{P[M}C^Q_{N]R}, \\
&=& \hat{R}_{MNP}{}^Q + 2\dd^Q_{[M}\hn_{N]}^{}\hn_P^{}\omega - 2\hg_{P[M}\hn_{N]}\hn^Q\omega \non\\
&& + 2\hn_{[M}^{}\omega\dd_{N]}^Q \hn_P\omega - 2\hn_{[M}^{}\omega\hg_{N]P}^{}\hn^Q\omega + 2\dd^Q_{[M}\hg_{N]P}^{} \left|\hn_R\omega\right|^2 \non.
\eea
The hatted quantities on the right hand side are constructed using the metric $\hg$, which is also used to raise and lower indices.  Contracting gives the Ricci tensor and scalar:
\bea
R_{MN} &=& \hat{R}_{MN} - \hg_{MN} \hn^2\omega + (D-2) \bigg(\hn_M\omega\hn_N\omega - \hn_M\hn_N\omega -\hg_{MN}\left|\hn_P\omega\right|^2\bigg), \\
R &=& e^{-2\omega}\left(\hat{R} - 2(D-1)\hn^2\omega - (D-2)(D-1)\left|\hn_M\omega\right|^2\right).
\eea
We will also extensively use the Laplacian expressed in terms of hatted variables for the metric~\C{conformalmetric}:
\be
\nabla^2 = e^{-2\omega}\left(\hn^2 + (D-2)\hn^M\omega\hn_M \right).
\ee
This is a special case of the formula
\be\label{2covder}
\nabla_M\nabla_N \Phi = \hat{\nabla}_M\hat{\nabla}_N \Phi - 2 \delta^P_{(M}\hat{\nabla}_{N)}\omega \hat{\nabla}_P\Phi + \hat{g}_{MN} \hat{\nabla}^P \omega \hat{\nabla}_P\Phi,
\ee
for a scalar field $\Phi$. For completeness, recall that the covariant derivative on a covariant vector field $A_M$ is written in terms of the Christoffel symbols as follows:
\be
\nabla_M A_N = \p_M A_N - \Gamma^P_{MN}A_P.
\ee
Also useful is the formula:
\bea \label{curvaturesquared}
e^{2\omega}\tr\left|R\right|^2_{MN} &=&\tr|\hat{R}|^2_{MN} + 4\hat{R}_{MPNQ}\left( \hat{\nabla}^P\omega \hat{\nabla}^Q\omega - \hat{\nabla}^P\hat{\nabla}^Q\omega \right) \\
&& +4\hat{R}_{(M}^{\phantom{(M}P}\hat{\nabla}_{N)}\omega\hat{\nabla}_P\omega - 4\hat{R}_{(M}^{\phantom{(M}P}\hat{\nabla}_{N)}\hat{\nabla}_P\omega \non\\ 
&& - 4 \hat{R}_{MN} \left|\hat{\nabla}\omega\right|^2 +\left(4-2D\right)\left|\hat{\nabla}\omega\right|^2\left(\hat{\nabla}_M\omega\hat{\nabla}_N\omega - \hat{\nabla}_M\hat{\nabla}_N\omega\right) \non\\
&& - 4\left|\hat{\nabla}\omega\right|^2\hat{\nabla}_M\hat{\nabla}_N\omega \non\\
&& + 2\left(8-2D\right)\left( \hat{\nabla}_P\omega\hat{\nabla}_{(M}\omega\hat{\nabla}_{N)}\hat{\nabla}^P\omega -\hat{\nabla}_P\hat{\nabla}_{(M}\omega\hat{\nabla}_{N)}\hat{\nabla}^P\omega \right) \non\\
&& - 4 \hat{\nabla}^2\omega \left(\hat{\nabla}_M\omega\hat{\nabla}_N\omega - \hat{\nabla}_M\hat{\nabla}_N\omega\right) \non\\
&&- 2\hat{g}_{MN} \left( \left(2-D\right) \left|\hat{\nabla}\omega\right|^4 + 2\hat{\nabla}_P\omega\hat{\nabla}_Q\omega\hat{\nabla}^P\hat{\nabla}^Q\omega - \hat{\nabla}_P\hat{\nabla}_Q\omega\hat{\nabla}^P\hat{\nabla}^Q\omega - 2\hat{\nabla}^2\omega\left|\hat{\nabla}\omega\right|^2\right). \non
\eea

\subsection{Self-duality of torsionful curvatures}\label{SDtorsion}

For a constant spinor, the supersymmetry variation of the gravitino implies that $\Omega_-$ is (anti-)self-dual in its tangent space indices:
\be
\Omega^{ab}_- = {\eta_6 \over 2}\epsilon^{abcd}\Omega^{cd}_-.
\ee
This implies the same property for $R_-$,
\be
R^{ab}_- = {\eta_6 \over 2}\epsilon^{abcd}R^{cd}_-.
\ee
From the definition of the torsionful connection, we note that the two torsionful curvatures are related under exchange of indices:
\be
R_{-mnpq} = R_{+pqmn} - 2 \partial_{[p}H_{qmn]}.
\ee
The self-duality of $R_-$ in its first two indices can then be expressed as the self-duality of a two-form constructed from $R_+$ and $dH$. Defining
\be
R_{+mn} = \hlf R_{+mnpq}dy^p\wedge dy^q, \, \ dH_{mn} = \iota_n\iota_m dH = 2\partial_{[m}H_{npq]}dy^p \wedge dy^q,
\ee
we see that
\be
\ast \left( R_{+mn} - \hlf dH_{mn} \right) = \eta_6 \left(R_{+mn} - \hlf dH_{mn} \right).
\ee
Or stated another way,
\be
\ast R_{+mn} = \eta_6 R_{+mn} -\hlf \eta_6 dH_{mn} + \hlf \ast dH_{mn}.
\ee
From this relation,  we see that $R_+$ is not self-dual as a two-form unless $dH_{mn} = 0$. One implication is that $\tr |R_+|^2$ and $\tr R_+\wedge R_+$ differ by terms proportional to $dH$.

Explicitly,
\bea
\tr R_+ \wedge \ast R_+ &\equiv & R_{+mn}\wedge \ast R_+^{mn} \cr
 &=& \eta_6 R_{+mn} \wedge R_+^{mn} -\hlf \eta_6 R_{+mn} \wedge dH^{mn} + \hlf R_{+mn}\wedge \ast dH^{mn}.
 \eea
We can simplify this further by noting that the first term in the last line is $\eta_6 \tr R_+ \wedge R_+$. Furthermore, in the last term, we use the property that for forms of the same degree $\omega \wedge \ast \eta = \eta \wedge \ast \omega$:
\bea
\tr R_+ \wedge \ast R_+  &=& \eta_6\tr R_+ \wedge R_+ - \hlf\eta_6 R_{+mn} \wedge dH^{mn}  + \hlf dH^{mn} \wedge \ast R_{+mn} \cr
&=&    \eta_6  \tr R_+ \wedge R_+ -{1 \over 4} \eta_6 dH_{mn} \wedge dH^{mn} +{1\over 4} dH_{mn}\wedge \ast dH^{mn}
\eea
In the supersymmetric case, 
\be
dH = {1 \over 3!}\partial_{[m}H_{npq]} dy^mdy^ndy^pdy^q = 2\ast \nabla^2 \Phi = {2\nabla^2\Phi \over 4!} \epsilon_{mnpq}dy^mdy^ndy^pdy^q.
\ee
Therefore, the two-form $dH_{mn}$ is
\be
dH_{mn} = \nabla^2\Phi \epsilon_{mnpq}dy^pdy^q,
\ee
with Hodge dual:
\be
\ast dH_{mn} = 2 \nabla^2\Phi g_{mp}g_{nq} dy^pdy^q.
\ee
This leads to the relations, 
\bea
dH_{mn} \wedge dH^{mn} &=& 0, \cr
dH_{mn}\wedge \ast dH^{mn} &=&48 \ast \left(\nabla^2 \Phi\right)^2 .
\eea
Putting these calculations together gives, 
\be
\tr R_+ \wedge \ast R_+ = \eta_6 \tr R_+ \wedge R_+  + 12 \ast  \left(\nabla^2 \Phi\right)^2.
\ee

\section{Connections and Curvatures}\label{curvatures}

In this appendix, we will summarize some formulae for connections and curvatures for the metric~\C{ns15metric}, excluding the $T^4$ factor. Define an orthonormal frame via,
\be\label{ortho}
ds^2 = - (e^0)^2 + (e^1)^2+ (e^r)^2  + (e^\th)^2+ (e^\phi)^2+(e^\psi)^2,
 \ee
 with 
 \bea
 & e^0 = {1\over \sqrt{f_1}} dx_0, \quad  e^1 = {1\over \sqrt{f_1}} dx_1, \quad e^r=\sqrt{f_5} dr, & \\ & e^\th = r\sqrt{f_5}\sin\psi d\th, \quad e^\phi = r\sqrt{f_5} \sin\psi \sin\th d\phi , \quad e^\psi = r\sqrt{f_5}  d\psi. &
 \eea
For this parametrization of $S^3$, both $\th$ and  $\psi$ run from $0$ to $\pi$, while $\phi$ runs from 0 to $2\pi$. We define the Hodge star acting on an orthonormal basis via,
\be
\ast \left(e^{a_1} \wedge \ldots \wedge e^{a_r} \right)= {1 \over (d-r)!} \e^{a_1 \ldots a_r}_{\phantom{a_1 \ldots a_r}a_{r+1} \ldots a_d} e^{a_{r+1}} \wedge \ldots \wedge e^{a_d},
\ee
where we take the convention that $\e_{01\ldots d-1} = 1$.

Using this frame, we can compute the spin connection with components:
 \bea
& \omega^{ab} = - \omega^{ba} & \\
& \omega^{0}_{\phantom{0}r} = {Q_1\over r^3 f_1  \sqrt{f_5} }e^0 , \quad  \omega^{1}_{\phantom{1}r} = {Q_1\over r^3 f_1  \sqrt{f_5} }e^1,  & \\ & \omega^{\th}_{\phantom{\th}r} = \left[ {1\over r \sqrt{f_5}} - {Q_5\over r^3 (f_5)^{3/2}} \right]e^\th,  \quad  \omega^{\th}_{\phantom{\th} \psi} = {1\over r\sqrt{f_5}}\cot{\psi}\, e^\th, \quad  \omega^{\phi}_{\phantom{\phi}r} = \left[ {1\over r \sqrt{f_5}} - {Q_5\over r^3 (f_5)^{3/2}} \right]e^\phi, & \\ &  \omega^{\phi}_{\phantom{\phi} \psi} = {1\over r\sqrt{f_5}} \cot\psi \, e^\phi, \quad  \omega^{\phi}_{\phantom{\phi} \th} = {1\over r\sqrt{f_5}}{\cot{\th} \over \sin\psi} \, e^\phi, \quad  \omega^{\psi}_{\phantom{\psi} r} = \left[ {1\over r \sqrt{f_5}} - {Q_5\over r^3 (f_5)^{3/2}} \right]e^\psi .&
 \eea
Note that all components of the spin connection vanish in the large $Q_1, Q_5$ limit. Combining the spin connection with $H$ gives $\omega_+$ which has components:
 \bea
& (\omega_+)^{0}_{\phantom{0}r} = {Q_1\over r^3 f_1  \sqrt{f_5} }e^0 , \quad (\omega_+)^{1}_{\phantom{1}r} = {Q_1\over r^3 f_1  \sqrt{f_5} }e^1,  & \\ & \omega^{\th}_{\phantom{\th}r} = \left[ {1\over r \sqrt{f_5}} - {Q_5\over r^3 (f_5)^{3/2}} \right]e^\th,  \quad  \omega^{\th}_{\phantom{\th} \psi} = {1\over r\sqrt{f_5}}\cot{\psi}\, e^\th, \quad  \omega^{\phi}_{\phantom{\phi}r} = \left[ {1\over r \sqrt{f_5}} - {Q_5\over r^3 (f_5)^{3/2}} \right]e^\phi, & \\ &  \omega^{\phi}_{\phantom{\phi} \psi} = {1\over r\sqrt{f_5}} \cos\psi \, e^\phi, \quad  \omega^{\phi}_{\phantom{\phi} \th} = {1\over r\sqrt{f_5}}{\cot{\th} \over \sin\psi} \, e^\phi, \quad  \omega^{\psi}_{\phantom{\psi} r} = \left[ {1\over r \sqrt{f_5}} - {Q_5\over r^3 (f_5)^{3/2}} \right]e^\psi .&
 \eea
The connection $\omega^{ab}$ can conveniently be written as a $6\times 6$ matrix of $1$-forms using the index ordering $(x_0, x_1, r, \th, \phi,\psi)$:
\be
\omega^{ab} = 
\bma
0 & 0 & h_1e^0 & 0 & 0 & 0 \\
0 & 0 & h_1e^1 & 0 & 0 & 0 \\
-h_1e^0 & -h_1e^1 & 0  & -h_2h_3e^{\th} & -h_2h_3e^{\phi} & -h_2h_3e^{\psi} \\
0 & 0 & h_2h_3e^{\th} & 0 & -{h_2 \cot \th \over \sin \psi }e^{\phi} & h_2\cot\psi e^{\th} \\
0 & 0 & h_2h_3e^{\phi}& {h_2 \cot \th \over \sin \psi }e^{\phi} & 0 & h_2\cot\psi e^{\phi} \\ 
0 & 0 &  h_2h_3e^{\psi}& -h_2\cot\psi e^{\th} & -h_2\cot\psi e^{\phi} & 0 
\ema.
\ee
Here we have defined radial functions,
\be\label{radialfunctions}
h_1(r) = { Q_1 \over r^3 f_1 \sqrt{f_5} }, \quad \ h_2(r) = {1 \over r \sqrt{f_5}}, \quad \ h_3(r) = 1 - {Q_5 \over r^2 f_5}.
\ee
We can express $H$ in terms of these functions, 
\be
{1\over 2} H = h_2(1-h_3)e^{\th}\wedge e^{\phi} \wedge e^{\psi} - h_1 e^0 \wedge e^1 \wedge e^r,
\ee
with the components of $H$ determined using:
\be
H = {1\over 3!} H_{abc} e^a e^b e^c. 
\ee  
To compute the torsional connection~\C{repeattorsionconnection}, we need a connection constructed from the components of $H$. This connection can also be conveniently expressed in matrix form:

\be
{1\over 2} H^{ab} = 
\bma
0 & h_1 e^r & - h_1e^1 & 0 & 0 & 0 \\
-h_1 e^r & 0 & -h_1e^0 & 0 & 0 & 0\\ 
h_1 e^1 & h_1e^0 & 0 & 0 & 0 & 0 \\
0 & 0 & 0 & 0 & h_2(1-h_3)e^{\psi} & -h_2(1-h_3)e^{\phi} \\
0 & 0 & 0 & -h_2(1-h_3)e^{\psi} & 0 & h_2(1-h_3)e^{\th} \\
0 & 0 & 0 & h_2(1-h_3)e^{\phi} & -h_2(1-h_3)e^{\th} & 0 
\ema. \non
\ee
The torsionful connection is given by the combination, 
\be
(\omega_+)^{ab}  = \resizebox{.9\hsize}{!}{$
\bma
0 & h_1e^r & h_1(e^0-e^1) & 0 & 0 & 0 \\
-h_1e^r & 0 & -h_1(e^0-e^1) & 0 & 0 & 0 \\
-h_1(e^0-e^1) & h_1(e^0 - e^1) & 0  & -h_2h_3e^{\th} & -h_2h_3e^{\phi} & -h_2h_3e^{\psi} \\
0 & 0 & h_2h_3e^{\th} & 0 & -{h_2 \cot \th \over \sin \psi }e^{\phi} + h_2(1-h_3)e^{\psi} & h_2\cot\psi e^{\th} - h_2(1-h_3)e^{\phi} \\
0 & 0 & h_2h_3e^{\phi}& {h_2 \cot \th \over \sin \psi }e^{\phi} - h_2(1-h_3)e^{\psi} & 0 & h_2\cot\psi e^{\phi} + h_2(1-h_3)e^{\th} \\ 
0 & 0 &  h_2h_3e^{\psi}& -h_2\cot\psi e^{\th}+ h_2(1-h_3)e^{\phi} & -h_2\cot\psi e^{\phi} - h_2(1-h_3)e^{\th} & 0 
\ema. $} \non
\ee
Using  a symbolic logic package, we find that the Pontryagin class associated to this torsionful connection is remarkably simple:
\be\label{gravcontribution}
 \tr\left(R_+ \wedge R_+\right) =  \frac{16 Q_5^2  \sin \psi
   \left(f_5(\sin \theta \sin \psi - \cos \psi) + 3 \sin\theta\sin \psi\right)}{r^5 f_5^4}dr \wedge d\th \wedge d \phi \wedge d\psi.
\ee
Note there is only support along the $\mathbb{R}^4$ transverse to the NS5-branes, and no dependence on the number of F1 strings, $Q_1$.  Perhaps more surprisingly, the square of the Riemann tensor, $|R_+|^2$, also does not depend on $Q_1$. This is a beautiful simplification that we will use in the main text.

\newpage
\ifx\undefined\bysame
\newcommand{\bysame}{\leavevmode\hbox to3em{\hrulefill}\,}
\fi

\end{document}